\documentclass[aps,prl,reprint,superscriptaddress]{revtex4-2}
\usepackage{amsfonts}
\usepackage{amsmath}
\usepackage{amssymb}
\usepackage{graphicx}
\usepackage{color}
\usepackage{amsmath}
\usepackage{float}
\usepackage{hyperref}

\newcommand \ra {a_j}
\newcommand \ia {b_j}
\newcommand \rz {c_j}
\newcommand \iz {d_j}

\newcommand \iu {\mathrm{i}}
\newcommand \im {\mathrm{i}}

\newcommand \kj {\mathbf{k}_j}

\newcommand \Lap {\nabla^2}

\newcommand \Grad {\vec{\nabla}}
\newcommand \nline {\nonumber \\}

\newcommand \be {\begin{eqnarray}}
\newcommand \ee {\end{eqnarray}}
\newcommand \ben {\begin{eqnarray}}
\newcommand \een {\end{eqnarray}}

\begin{document}

\title{Mesoscale Defect Motion in Binary Systems: Effects of Compositional
Strain and Cottrell Atmospheres}

\begin{abstract}
The velocity of dislocations is derived analytically to incorporate 
and predict the intriguing effects induced by the preferential solute segregation 
and Cottrell atmospheres in both two-dimensional and three-dimensional binary 
systems of various crystalline symmetries. The corresponding mesoscopic 
description of defect dynamics is constructed through the amplitude
formulation of the phase-field crystal model which has been shown to accurately
capture elasticity and plasticity in a wide variety of systems. Modifications 
of the Peach-Koehler force as a result of solute concentration variations
and compositional stresses are presented, leading to interesting new
predictions of defect motion due to effects of Cottrell atmospheres.  
These include the deflection of dislocation glide paths, the variation of 
climb speed and direction, and the change or prevention of defect annihilation, 
all of which play an important role in determining the fundamental behaviors of complex defect network and dynamics. The analytic results are verified by numerical simulations. 
\end{abstract}

\author{Marco Salvalaglio}
\email{marco.salvalaglio@tu-dresden.de} 

\author{Axel Voigt}
\affiliation{Institute  of Scientific Computing,  TU Dresden,  01062  Dresden,  Germany} 
\affiliation{Dresden Center for Computational Materials Science, TU Dresden,  01062  Dresden,  Germany} 

\author{Zhi-Feng Huang}
\affiliation{Department of Physics and Astronomy, Wayne State University, Detroit, Michigan 48201, USA}

\author{Ken R. Elder}
\affiliation{Department of Physics, Oakland University, Rochester, Michigan 48309, USA}


\maketitle

In crystalline systems, topological defects, such as dislocations and grain 
boundaries, play a significant role in controlling system properties. 
For example, in polycrystals the average grain size plays a major role in 
determining the magnitude of the  magnetic coercivity \cite{Herzer2013,Xue2008}, 
yield stress \cite{Yip1998,Petch1953}, and thermal conductivity \cite{Fan2017}. 
It is, thus, of critical importance to understand the nature 
of defect motion and the corresponding elastoplastic mechanisms during 
the evolution of nonequilibrium material systems, which control, e.g., 
grain coarsening rates and, hence, the resulting defected structures and 
configurations of polycrystalline systems. Dislocations lead to 
strains in crystalline lattices which, in turn, are offset to some extent 
in binary alloys by phase segregation, or \textit{Cottrell atmospheres}
\cite{Cottrell1949,Cottrell1949_2,CottrellBOOK} near the dislocation cores.
This segregation influences the motion of dislocations and grain
boundaries \cite{LUCKE1957,CAHN1962,LUCKE1971,HILLERT1976,HILLERT2004} by 
modifying the effective Peach-Koehler driving force that acts on the 
dislocations. Typically, this phenomenon was investigated by focusing 
on concentration profiles and stress distribution around dislocations 
\cite{Cahn2013,Hirth2014,*Cahn2014,*Hirth2014b,MISHIN2016,HIRTH2017}
and the force-velocity curves for defect motion.
In most cases, either continuum modeling of defect motion or atomistic
description was considered. This also applies to computational studies, 
from the first numerical approaches tracking concentration profiles and 
velocities \cite{Yoshinaga1971,Takeuchi1979} up to the most recent advanced 
numerical investigations accounting for segregation at both dislocations
\cite{Zhang_2008,Sills2016,Gu2020} and grain boundaries 
\cite{MISHIN2019,Koju2020DirectAM,DarvishiKamachali2020}. 

Given the complex, mesoscopic characteristics of the defect motion, it is of
fundamental importance to bridge the above two ends of the description spectrum
at atomistic and long-wavelength continuum scales and examine the key features 
of mesoscale effects \cite{Rollett2015}. This often requires coarse-grained
approaches, handling large length scales through continuum density fields 
that still retain relevant microscopic details of the atomic structures of 
defects. Although much progress has been made on this front, such as 
those based on the multiscale  phase-field crystal (PFC) method
\cite{Elder2002,Elder2004,Elder2007}, most studies have been focused on 
the defect dynamics in single-component systems
\cite{Skaugen2018b,Skaugen2018,SalvalaglioNPJ2019,SalvalaglioJMPS2020}, 
while the understanding of the defect behavior in alloys or multicomponent 
systems, especially the novel elastoplastic properties originated from
the coupling to compositionally generated effects, is still limited. 

In this Letter, we construct a mesoscopic description of dislocation 
dynamics for binary alloy systems, through an analytic formulation of 
dislocation velocities as a function of the solute expansion coefficient 
and alloy concentration (i.e., compositional strain), for various 
two-dimensional (2D) and three-dimensional (3D) crystalline symmetries. 
It is based on the PFC model in its complex amplitude expansion formalism 
(APFC) \cite{Goldenfeld2005,Athreya2006,ElderPRE2010,Huang2010,SalvalaglioNPJ2019} and extends the current description of defect velocities in 2D single-component 
systems for triangular lattices \cite{Skaugen2018b} to incorporate 
the key effects induced by local concentration variations around 
defects in both 2D and 3D binary systems. 
The mesoscale character of this framework results from 
the coarse-grained description of the lattice structure, deformation fields,
and variations of the alloy concentration conveyed by the APFC model, although dislocations are still described as individual objects rather 
than through an averaged dislocation density.
Numerical APFC simulations are used to verify the analytic calculations, 
illuminating the solute preferential segregation at defects and, importantly, its influence on defect motion for different configurations and crystal 
symmetries. One of the intriguing results is the prediction of 
the deflection of dislocations from the glide paths and the change of
climb direction that would be followed in a pure system. This could even 
prevent defect annihilation, indicating the novel effect of Cottrell 
atmospheres and the compositionally induced stress on defect dynamics.

The original binary PFC model \cite{Elder2007} is formulated in terms of 
the dimensionless atomic number density variation field $\varrho(\vec{r},t)$ 
and a solute concentration field $\psi(\vec{r},t)$.  For the purposes 
of this work it is useful to consider the corresponding amplitude expansion
representation \cite{ElderPRE2010} in which the density field is expanded by 
\begin{equation} \varrho =  \sum_n
\eta_{n}e^{\iu \vec{q}_{n} \cdot \vec{r}} + {\rm c.c.},
\end{equation}
where $\eta_{n}(\vec{r},t)$ are complex, slowly varying amplitudes,
the wave vectors $\vec{q}_{n}$ specify a given crystalline symmetry,
``${\rm c.c.}$" represents the complex conjugate, and, for simplicity, 
the average of $\varrho$ is set as constant and zero. 
By assuming the lattice spacing $R$ to be linearly proportional to $\psi$
(Vegard's law), we have $R=R_0(1+\alpha\psi)$ with $\alpha$ the solute expansion
coefficient. The dynamic equations for $n$ and $\psi$ in dimensionless form
are written as \cite{free} 
\begin{equation}
\frac{\partial \eta_{n}}{\partial t} 
= -{q}_{n}^2\frac{\partial F}{\partial \eta^*_{n}}, \qquad 
\frac{\partial \psi}{\partial t} = \nabla^2 \frac{\partial F}{\partial \psi},
\label{eq:evo}
\end{equation}
respectively, where
\begin{eqnarray}
F  = \int \bigg[&&
\frac{\Delta B_0}{2} \Phi+\frac{3v}{4}\Phi^2  +\sum_{n}\left(B_0^x|{\cal
G}_n\eta_n|^2 -\frac{3v}{2}|\eta_n|^4\right) \nline
&&+f^{\rm s}(\{\eta_n\})+
(\omega+B_2^l\Phi)\frac{\psi^2}{2}+\frac{u}{4}\psi^4 \nline &&
-2B_0^x\alpha\sum_nq_n^2\left(\eta_n {\cal G}_n^* \eta_n^*+{\rm
c.c.}\right)\psi
\bigg]d\vec{r},
\label{eq:energy}
\end{eqnarray}
$\Phi=2\sum_n|\eta_n|^2$, ${\cal G}_n = \Lap+2\iu\vec{q}_n\cdot \Grad$, 
and $\Delta B_0$, $v$, $B_0^x$, $w$, $B_2^l$, and $u$ are model 
parameters as described in Ref.~\cite{Elder2007}. Here $f^{\rm s}(\{\eta_n\})$ is a polynomial in $\eta_n$ (and $\eta_n^*$) that depends on the specific
crystalline symmetry under consideration (see Supplemental Material
\cite{Suppl}). It can be shown that, given
$\vec{q}_n$ the basic wave vectors corresponding to a pure system, 
the equilibrium wave vectors for binary systems read 
$\vec{q}_n^{\,\rm eq}=\vec{q}_n\sqrt{1-2\alpha\psi}$ \cite{Huang2010}. 
This amplitude model as written does not impose instantaneous mechanical 
equilibrium, nor does it contain Peierls barriers to defect motion, 
although both effects have been included in more complex models \cite{SalvalaglioJMPS2020,Skaugen2018,Huang2013,*Huang2016}.

A dislocation in a crystalline lattice corresponds to a 
discontinuity in the phase ($\theta_n$) of the complex 
amplitudes which can be written as $\eta_n = \phi_n 
e^{\iu\theta_n}$.  The discontinuity in the phase 
corresponds to a discontinuity in the displacement 
field $\vec{u}$ that enters continuum elasticity 
theory, since this displacement is equivalent to 
setting $\theta_n = -\vec{q}_n \cdot \vec{u}$ \cite{ElderPRE2010,Heinonen2014}.
More explicitly, a dislocation with Burgers vector $\vec{b}$ is defined by
$\oint d\vec{u} = \vec{b}$, corresponding to $\oint d\theta_n = -\vec{q}_n
\cdot \vec{b} = -2\pi s_n$, where $s_n$ is the winding number.  As in 
Ref.~\cite{Skaugen2018b}, in what follows the vortex solution $\eta_n \propto
x - i s_n y $ will be considered with $s_n = \pm 1$.

To examine the influence of solute concentration on dislocation motion 
it is useful to define the Burgers vector density $\vec{B}(\vec{r}$) as
$\vec{B}(\vec{r}) = \sum_m \vec{b}_m \delta(\vec{r}-\vec{r}_m)$,
where $\vec{b}_m$ and $\vec{r}_m$ are the Burgers vector and position 
of the $m$th dislocation, respectively. At a dislocation core some of 
the amplitudes go to zero; it is, thus, useful to make a transformation 
from spatial coordinates to the real and imaginary components of the 
complex amplitudes. Generalizing Ref.~\cite{Skaugen2018b} to the case
of a point dislocation in 2D or an edge dislocation in 3D, the
transformation leads to 
\begin{equation}
\vec{B}=- \beta \sum_n \vec{q}_n D_n \delta(\eta_n),
\quad  D_n = \frac{\varepsilon_{jk}}{2i}\partial_j 
\eta_n \partial_k \eta_n^*,
\end{equation}
where $\beta=2\pi/\sum(q^n_j)^2$ for $j=x,y,z$, 
$\varepsilon_{jk}$ is the Levi-Civita symbol, and the Einstein
summation convention is implied. By writing $\vec{B}$ in terms of 
the amplitudes the dynamics of $\vec{B}$ is determined by
\begin{equation}
\frac{\partial B_i }{\partial t} = - \partial_j {\cal J}_{ij} 
= - \partial_j
\bigg[ \sum_m b_i^m v_j^\alpha \delta(\vec{r}-\vec{r}_m) \bigg],
\end{equation}
with the dislocation velocity
\begin{equation}
v_j^m = \frac{\beta}{2\pi} 
\sum_n  \frac{(\vec{q}_n \cdot \vec{b}_m)^2}
{|\vec{b}_m|^2}\frac{J_j^n}{D_n}, \ \ 
J^n_j = \varepsilon_{jk}{\rm Im}(\dot{\eta}_n \partial_k \eta_n^*).
\label{eq:Jnj}
\end{equation}
Near the dislocation core the dynamic equation 
of motion for $\eta_n$ can be approximated as
\begin{equation}
\frac{\partial \eta_n}{\partial t} 
= -{q}_n^2 B_0^x \left[{\cal G}_n^2 \eta_n - 2\alpha q_n^2
\left(\psi {\cal G}_n \eta_n + {\cal G}_n \eta_n \psi
\right)
\right],
\end{equation}
which can be further simplified to 
\begin{equation}
\frac{\partial \eta_n}{\partial t} = -\iu 8{q}_n
^2 B_0^x  \vec{q}_n \cdot \Grad \phi_n
\left(\vec{q}_n \cdot \Grad \theta_n + q_n^2 \alpha \delta \psi
\right) e^{\iu \theta_n},
\label{eq:etat}
\end{equation}
with $\delta \psi = \psi - \psi_{\rm core} \approx \bar{\psi} - \psi_{\rm core}$, 
i.e., the difference between the concentration far away from the dislocation
and its value at the defect core, where $\bar{\psi}$ is the average
concentration. Substituting Eq.~\eqref{eq:etat} into Eq.~\eqref{eq:Jnj} 
and using the results 
$\iu\partial_j\eta_n= - 1/s_n \varepsilon_{jk}\partial_k \eta_n$ and
${\rm Im}(\partial_j \eta_n \partial_k \eta_n^*) = \varepsilon_{jk}D_n$ \cite{Skaugen2018b} 
leads to 
\begin{equation}
\frac{J_j^n}{D_n} = \frac{8}{s_n} {q}_n^2 B_0^x 
\varepsilon_{jk}q_k^n\left(q_l^n q_p^n \partial_l u_p - q_n^2
\alpha \delta\psi \right),
\end{equation}
and in turn, 
\begin{equation}
\!\!v_j^m = \frac{8 \beta B_0^x b^m_i}{|\vec{b}_m|^2} 
\varepsilon_{jk}
\sum_n q_n^2 q^n_i q^n_k \left(q^n_l q^n_p \partial_l u_p - 
q_n^2 \alpha \delta\psi\right).
\label{eq:vj1}
\end{equation}
Furthermore, since Eq.~\eqref{eq:vj1} is symmetric in 
$l$ and $p$ it can be written in terms of the strain 
tensor $U_{ij}=(\partial_i u_j + \partial_j u_i)/2$ as 
follows,
\begin{equation}
\!v_j^m = \frac{8 \beta B_0^x b_i^m}{|\vec{b}_m|^2} 
\varepsilon_{jk} \sum_n q_n^2 q^n_i q^n_k \left( q^n_l q^n_p U_{lp} 
- q_n^2 \alpha \delta\psi \right).
\label{eq:vj2}
\end{equation}
Equation \eqref{eq:vj2} is consistent with the classical Peach-Koehler force
\cite{pk2019}, since the corresponding stress ($\sigma_{ij})$ is proportional 
to the strain, i.e., $\sigma_{jk} = \lambda_{jklm} U_{lm}$, where $\lambda_{jklm}$ 
is the rank-four elastic modulus tensor \cite{LLEL}.  More explicitly, the 
calculations reported in Ref.~\cite{Skaugen2018b} can be easily extended to more
complex crystal structures where the magnitude $\phi_n$ of the complex amplitudes
$\eta_n$ are not all the same in equilibrium, giving 
\begin{equation}
\sigma_{jk} = 8 B_0^x U_{lp} \sum_n \phi_n^2 q^n_j q^n_k q^n_l q^n_p.
\label{eq:sigma_jk}
\end{equation}
Note that both Eqs.~\eqref{eq:vj2} and \eqref{eq:sigma_jk} 
are of mesoscopic nature, given the mesoscale amplitudes, displacements, and
concentration variations.
For the case of a 2D triangular lattice or a 3D bcc crystal where
it is possible to construct the lattice by retaining only one mode of
the lowest order (with $q_n=1$), the velocity takes the form
\begin{equation}
v_j^m = M \varepsilon_{jk}\left( \sigma_{ki} b_i^m 
- 4B_0^x\phi_0^2\alpha \delta \psi b_i^m \sum_n q^n_i q^n_k 
\right),
\label{eq:vj3}
\end{equation}
with a mobility $M= 2\beta/(\phi_0^2|\vec{b}_m|^2)$
and the equilibrium amplitude magnitude $\phi_0$ of the lowest-order mode. 
The last term in Eqs.~\eqref{eq:vj2} and \eqref{eq:vj3} accounts for the
new contribution from the compositionally generated stress, as a result of
the compositional strain ($\sim \alpha \psi$) arising from local concentration
variations particularly solute preferential segregation (Cottrell atmospheres)
around defects. Thus Eqs. \eqref{eq:vj2} and \eqref{eq:vj3} provide explicit
predictions for the influence of solute concentration on dislocation motion 
for general crystalline symmetries and are the main results of this Letter. 

\begin{figure}[b]
\includegraphics[width=0.49\textwidth]{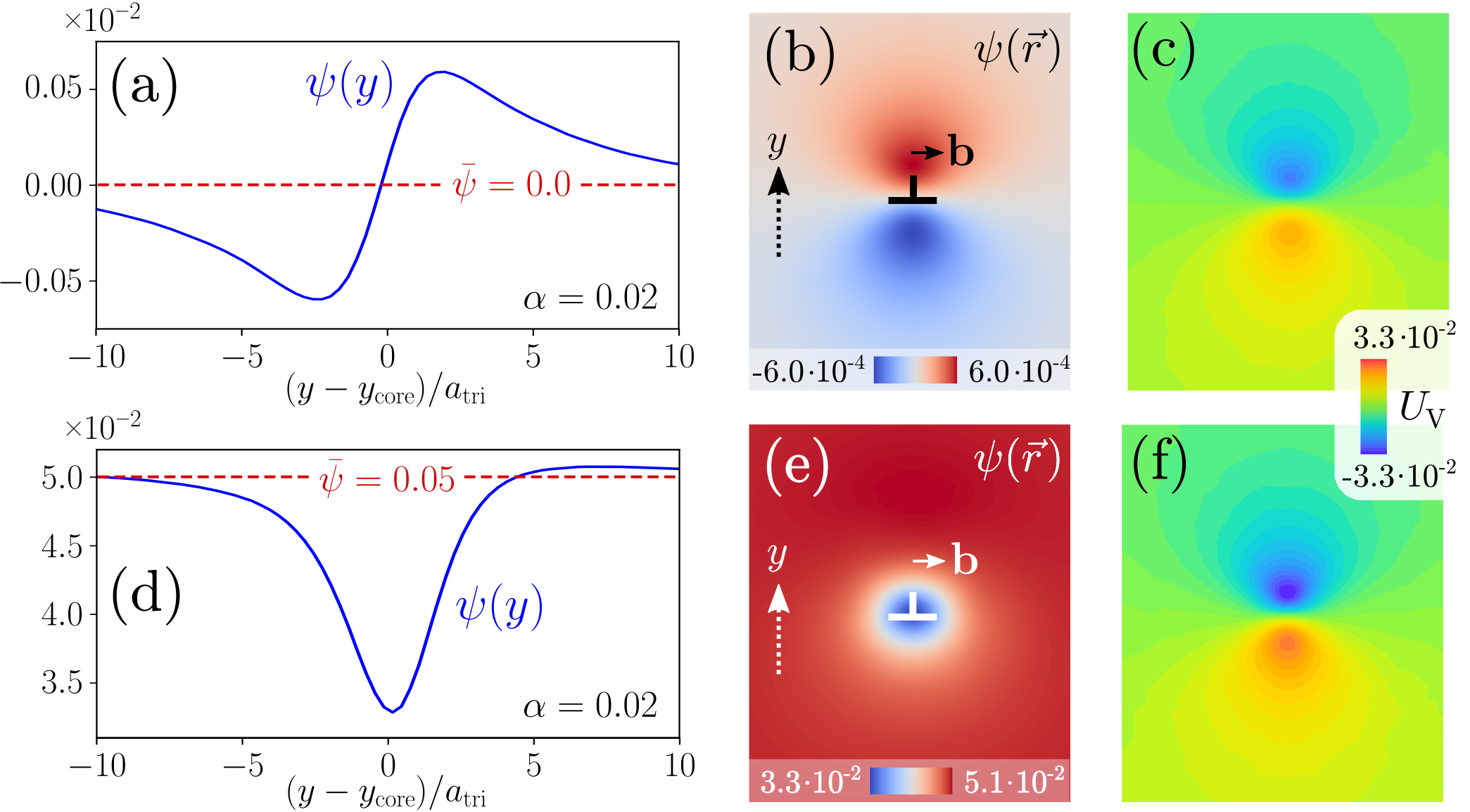}
\caption{Profiles of phase segregation and strain
around a dislocation in a 2D 
triangular crystal with $\vec{b}=(b_x,0)$: (a)-(c) $\psi(y-y_{\rm core})$,
$\psi(\vec{r})$, and $U_{\rm V}=U_{xx}+U_{yy}$ distributions
for $\alpha=0.02$ and $\bar{\psi}=0$;
(d)-(f) $\psi(y-y_{\rm core})$, $\psi(\vec{r})$ and 
$U_{\rm V}$ distributions for $\alpha=0.02$ and $\bar{\psi}=0.05$.}
\label{fig:figure1}
\end{figure}

In what follows we consider the lowest-order mode expansion that is a good
approximation of the full PFC models near melting and is exact for the APFC.
A 2D triangular ($T$) or honeycomb lattice requires three reciprocal
vectors, $\vec{q}_1$=$\langle-\sqrt{3}/2,-1/2\rangle$, 
$\vec{q}_2$=$\langle0,1\rangle$ and $\vec{q}_3$=$ -\vec{q}_1-\vec{q}_2$, and
thus,
\begin{equation}
\begin{split}
v_x=&\gamma\bigg[ 2 U_{xy} b_x + (U_{xx}+3U_{yy}  )b_y - 4\alpha \delta \psi b_y \bigg],\\
v_y=&-\gamma\bigg[ 2 U_{xy} b_y + (3U_{xx}+U_{yy}  )b_x -  4\alpha \delta \psi b_x \bigg],
\end{split}
\label{eq:vtri}
\end{equation}
where $\gamma\equiv 4\pi B_0^x/|\vec{b}_m|^2$.
Explicit expressions for a point dislocation in 2D square lattice and an 
edge dislocation in 3D bcc and fcc systems are given in Supplemental 
Material \cite{Suppl}.

To validate the above analytical results we numerically integrate the amplitude
Eqs.~\eqref{eq:evo} and \eqref{eq:energy} for some representative cases.
Here we consider the system in the single-phase regime of the phase diagram 
and do not investigate the influence of dislocations on phase separation
in a two-phase state \cite{DarvishiKamachali2020}.
The simulations exploit the finite element toolbox AMDiS 
\cite{Vey2007,Witkowski2015} and build on
the algorithms described in Refs.~\cite{SalvalaglioAPFC2017,Praetorius_2019}.
The initial concentration field is set
to be uniform, i.e., $\psi(\vec{r})=\bar{\psi}$. The initial conditions
for amplitudes are set to encode a distortion of a relaxed crystal having 
equilibrium wave vectors $\vec{q}_n^{\,\rm eq}$. 
Details are reported in the Supplemental Material \cite{Suppl}. 

\begin{figure}[b]
\includegraphics[width=0.48\textwidth]{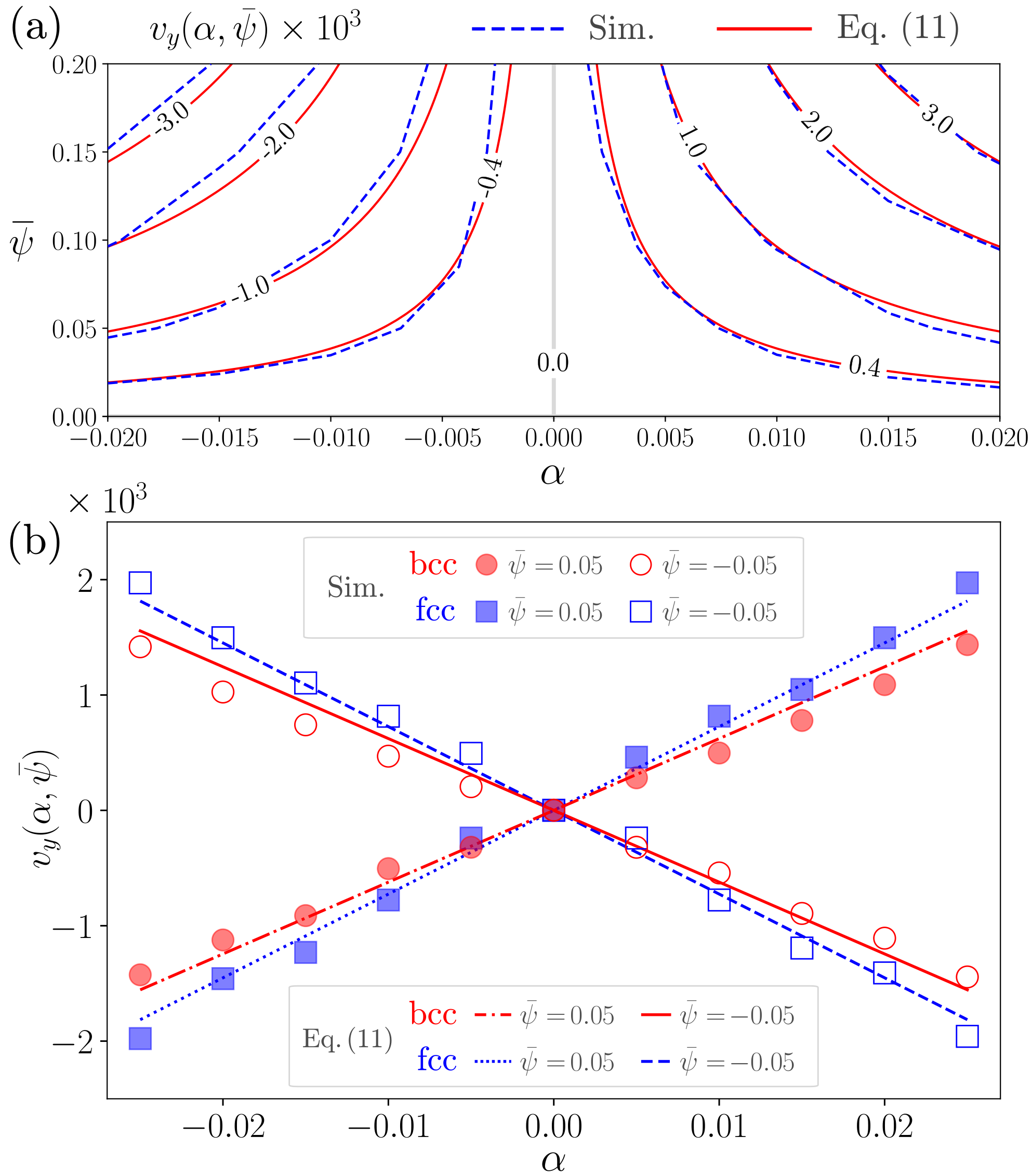} 
\caption{Segregation-induced dislocation velocity $v_y(\alpha,\bar{\psi})$
evaluated from APFC simulations and the analytic result Eq.~\eqref{eq:vj2},
for (a) triangular and (b) bcc and fcc symmetries.}
\label{fig:figure2}
\end{figure}

We first consider an edge dislocation in a 2D triangular lattice, with Burgers
vector $\vec{b}=(\pm b_x,0)$ and $b_x=a_{\rm tri}=4 \pi / \sqrt{3}$, forming
between regions with opposite deformation $u_x$ and corresponding to an
equilibrium configuration in a pure system where no motion is expected  
with zero Peach-Koehler force. 
The solute segregation near defect cores is illustrated in 
Fig.~\ref{fig:figure1} showing the $\psi(\vec{r})$ profiles computed. 
For $\bar{\psi}=0$ two lobes with positive and negative $\psi(\vec{r})$ 
form [see Figs.~\ref{fig:figure1}(a)--\ref{fig:figure1}(b)]. For
$\bar{\psi}\neq 0$ the concentration shows a well-shaped distribution
with a slightly asymmetric profile around the defect core
[Figs.~\ref{fig:figure1}(d)--\ref{fig:figure1}(e)].
The corresponding volumetric strain field $U_V=U_{xx}+U_{yy}$, 
which accounts for both lattice distortion and compositional strain (see 
Supplemental Material \cite{Suppl}), is reported in Figs.~\ref{fig:figure1}(c) 
and \ref{fig:figure1}(f). Notice that for $\bar{\psi}=0$ 
the segregation slightly opposes the lattice deformation induced by the defect. 
For $\bar{\psi}=0.05$, the solute depletion at the core is observed, while 
an asymmetric contribution is present that resembles the effect observed 
for $\bar{\psi}=0$.

As described by Eq.~\eqref{eq:vj2} or \eqref{eq:vj3}, the preferential 
segregation (i.e., Cottrell atmospheres) at dislocations affects the defect 
velocity, with quantitative effects depending on the lattice symmetry. 
For the triangular case, the velocities $v_y^{\rm T}(\alpha,\bar{\psi})$ 
obtained by simulations [Fig.~\ref{fig:figure2}(a), blue dashed lines] match well with the prediction of Eq.~\eqref{eq:vtri} [Fig.~\ref{fig:figure2}(a), 
red solid lines] with $\delta \psi$ extracted from simulations. 
Note that the velocity values in Fig.~\ref{fig:figure2} have been 
subtracted by a small correction $v_y(0,0)$. This small drift is caused
by the weak anisotropy in APFC Eq.~\eqref{eq:energy} for displacements 
with the same magnitude but different sign \cite{Huter2016}; It is not included 
in Eq.~\eqref{eq:vj3}, and is found to be independent of $\alpha$ and 
$\bar{\psi}$. The velocities of dislocations in bcc and fcc crystals, 
forming between layers with opposite deformations $u_x$ (see 
Supplemental Material \cite{Suppl}) are also calculated by both numerical 
simulations and Eq.~\eqref{eq:vj2}, showing a good agreement as well,
as demonstrated in Fig.~\ref{fig:figure2}(b). In these 3D cases we have
set the lattice displacements to obtain edge dislocations parallel to the 
$z$ axis and $\vec{b}^{\rm s}=b_x^{\rm s} \hat{x}^{\rm s}$, with $\hat{x}^{\rm B}=[100]$, $b_x^{\rm B}=2\pi\sqrt{2}$ (bcc) 
and $\hat{x}^{\rm F}=[110]$, $b_x^{\rm F}=\pi\sqrt{6}$ (fcc). 
The simulation results verify the linear dependence of dislocation velocity 
on the compositional strain or stress as predicted by Eq.~\eqref{eq:vj2}.

\begin{figure}[t]
\includegraphics[width=0.47\textwidth]{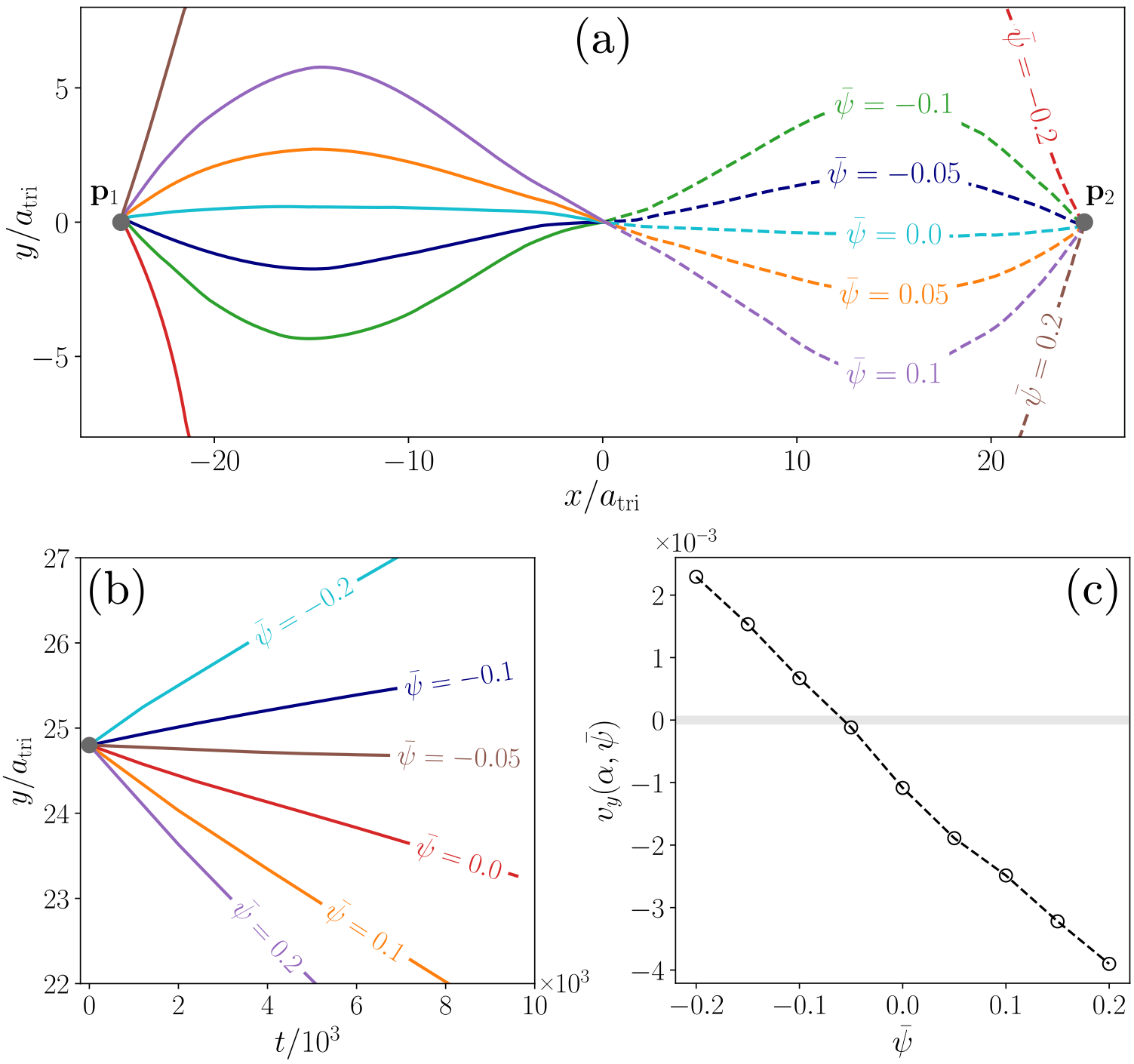} 
\caption{(a) Trajectories of two 2D edge dislocations in the G configuration,
for $\alpha=0.02$ and various values of $\bar{\psi}$, with $\bar{\psi}=0$
corresponding to pure glide.
(b) Time evolution of the $y$ position of the upper dislocation in
configuration C with $\alpha=0.02$. $\bar{\psi}=0$ corresponds to pure 
climb. (c) Dislocation velocity as a function of $\bar{\psi}$ 
as identified from (b).} 
\label{fig:figure3}
\end{figure}

More insights on the effects predicted by Eqs.~\eqref{eq:vj2} and \eqref{eq:vj3}
can be obtained by focusing on nonequilibrium configurations involving
defect dynamics of glide and climb. For instance, we consider dislocation pairs 
in a triangular lattice that are expected to move by pure \textit{glide} (G) and
\textit{climb} (C), with $\vec{b}_{1,2}=(\pm a_{\rm tri},0)$ and positions
$(\pm d,0)$ and $(0,\pm d)$, respectively, in a $L \times L$ simulation
box with $L \gg 2d$ and $d \sim 28 a_{\rm tri}$. These configurations are
initialized using the displacement field induced by straight edge 
dislocations \cite{anderson2017} and the corresponding $\eta_n$
\cite{SalvalaglioJMPS2020} (see Supplemental Material \cite{Suppl}). 
The dynamics of these defects, depending on $\alpha$ and $\bar{\psi}$, 
is illustrated in Fig.~\ref{fig:figure3}(a)--(c) for $\alpha=0.02$ and
different values of $\bar{\psi}$. For configuration G, a nonzero $v_y$ 
component is obtained, directly corresponding to the ones reported in 
Fig.~\ref{fig:figure2}(a), while a small but nonzero $v_x$ encodes the 
effect of strain induced by the presence of a second dislocation, reproducing 
the effect of the Peach-Koehler force that leads to defect annihilation 
by pure glide in single-component systems. Interestingly, at relatively large 
values of $\alpha\bar{\psi}$, the annihilation of the dislocations by glide 
can be avoided [see Fig.~\ref{fig:figure3}(a) and Supplemental Videos]. 
This new effect can be understood through Eq.~\eqref{eq:vtri}: Given
$|v_y^{t=0^+}|>|v_x^{t=0^+}|$, this absence of annihilation would occur
when $|\alpha \delta \psi| > |U_{xy}|/2$ with $U_{xy}$
corresponding to the strain field component caused by the other 
dislocation in the dipole while $U_{xx}^{t=0^+}=U_{yy}^{t=0^+}=0$. 
Therefore, as driven by purely thermodynamic driving forces, 
a threshold value exists for $|\alpha \delta \psi|$ above which the
defect annihilation is prevented, with dislocations moving away from
the traditional glide planes. 

For configuration C, the velocity is oriented only along the $y$ axis 
as predicted by Eq.~\eqref{eq:vtri} as $b_y=0$. The symmetry of the 
simulations setup is such that the two dislocations are separated by $L_y/4$ (leaving them a distance $>3L_y/4$ from their periodic counterpart). 
The contribution of compositional strain
can then accelerate, slow down, or even prevent the annihilation, 
as illustrated in Figs.~\ref{fig:figure3}(b)--\ref{fig:figure3}(c)
(see also Supplemental Videos).
A change of the sign of the dislocation velocity is shown in
Fig.~\ref{fig:figure3}(c), implying that the defects are moving toward
their farther away periodic counterpart. In this case, a threshold can 
be estimated through Eq.~\eqref{eq:vtri} again as the condition $v_y=0$, 
yielding $\alpha \delta \psi=(3U_{xx}+U_{yy})/4$, with $U_{xy}=0$.

 \begin{figure}[t]
\includegraphics[width=0.47\textwidth]{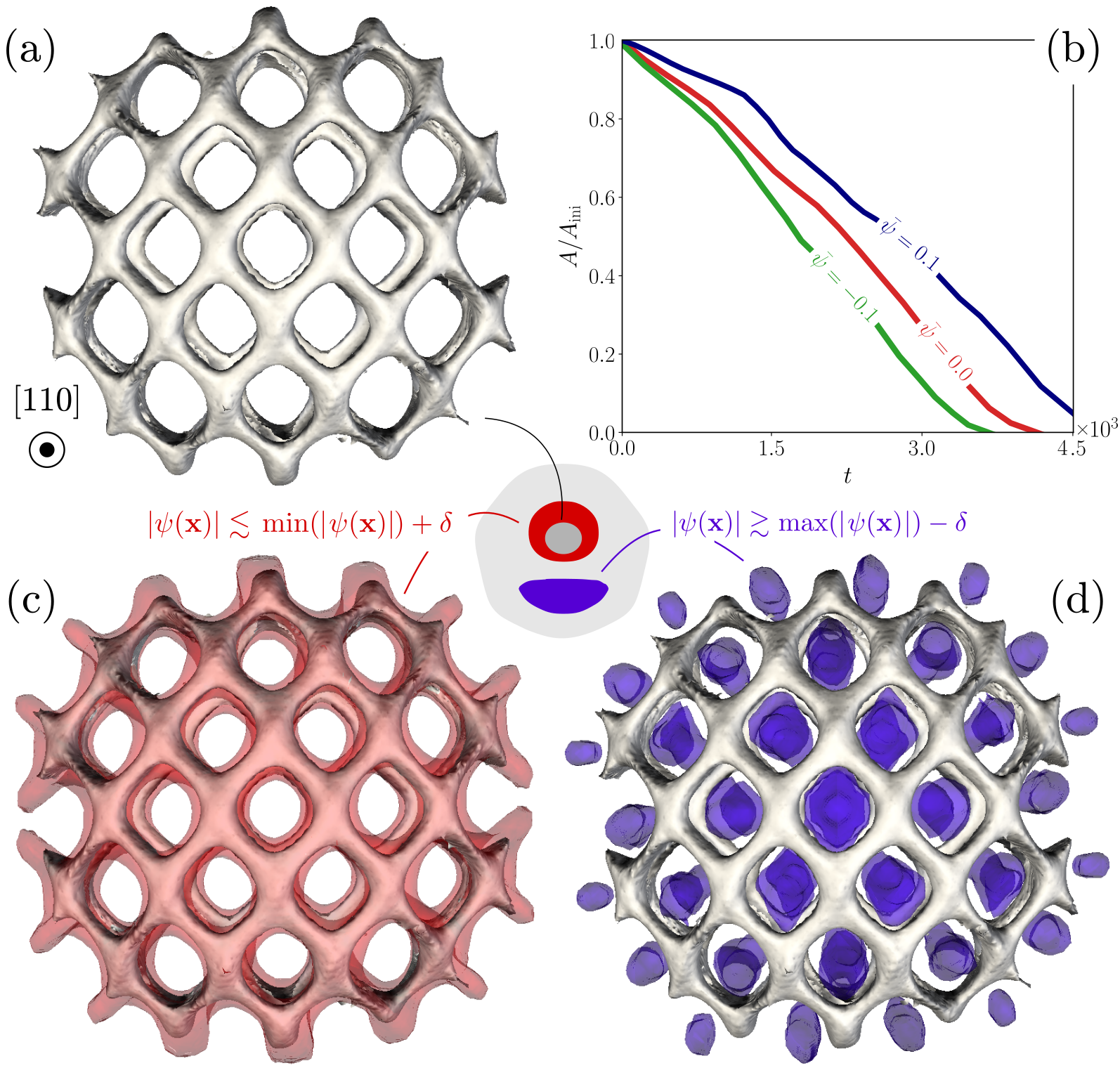} 
\caption{(a) A network of dislocations at the boundary of an 
inclusion rotated by 10$^\circ$ about the $[110]$ direction in a bcc 
crystal [Regions shown: $\Phi<0.85\max(\Phi)$]. (b) Normalized area of 
the grain boundary as a function of time, for $\alpha=0.02$. 
(c)--(d) Concentration segregation at defects 
for the dislocation network of panel (a). The middle inset shows 
spatial profiles of quantities of panels (a), (c), and (d) in a 
defect cross section.} 
\label{fig:figure4}
\end{figure}

A more complex configuration involving many defects is also examined, forming
the dislocation network embedded in a crystalline matrix. In particular we 
illustrate the case of a 3D bcc crystal with an embedded grain tilted by 
$10^{\circ}$ about the [110] direction and of radius 
$\sim 15 a_{\rm bcc}$ with $a_{\rm bcc}=2 \pi \sqrt{2}$
\cite{SalvalaglioPRM2018,SalvalaglioNPJ2019}. As illustrated in
Fig.~\ref{fig:figure4}(a), a spherical network of dislocations, namely 
a small-angle grain boundary, forms and it is expected to shrink 
anisotropically \cite{Doherty1997,Yamanaka2017,SalvalaglioPRM2018}. 
The simulated solute segregation at defects is illustrated in Figs.~\ref{fig:figure4}(c)--\ref{fig:figure4}(d). The rate of 
shrinkage of the dislocation network is
affected by the solute expansion coefficient and average concentration 
[see Fig.~\ref{fig:figure4}(b)], which can be ascribed to the interplay 
of changes of defect dynamics as reported in Fig.~\ref{fig:figure3}.
It is noted that analytic expressions \eqref{eq:vj2} 
and \eqref{eq:vj3} apply to straight dislocations. Extensions to arbitrarily 
curved dislocations in 3D, and, in turn, to configurations as in 
Fig.~\ref{fig:figure4}, are expected to follow by accounting for the 
local orientation of curved dislocation lines and the corresponding distortion in the lattice, e.g., through the use of the Nye tensor
\cite{nye1953some,hartley2005characterization}. 
This formalism is being developed for single-component systems and will 
be extended to binary alloys.

In conclusion, through a coarse-grained approach we have identified
analytic expressions for the velocities of dislocations in binary
systems. The results predict the effects of compositional stress
generated by the solute preferential segregation near the dislocation
cores (i.e., Cottrell atmospheres) for different 2D and 3D crystalline
symmetries, as confirmed by numerical simulations of the APFC
model. While the influence of solute concentration on the magnitude of
dislocation velocity was expected, this work also predicts some novel,
segregation-induced behaviors of defect dynamics, such as the
velocity components parallel to the Burgers vector in glide, leading to
deflections from the traditional glide planes that could avert
defect annihilation, as well as altering of dislocation climb
rate, reverse of climb direction, or even stagnation. 
The mesoscopic formulation constructed here provides a powerful tool 
to understand the nature of defect motion in binary alloys which 
controls the structural dynamics and properties of the material system.

\begin{acknowledgments}
We gratefully acknowledge the computing time granted by J\"ulich Supercomputing
Centre (JSC) within Project No. HDR06, and by ZIH at TU Dresden. 
M.S. acknowledges support from the Emmy Noether Programme of the German Research Foundation (DFG) under Grant No. SA4032/2-1. A.V. acknowledges support from the German Research Foundation (DFG) within SPP1959 under Grant No. VO899/19-2. K.R.E. acknowledges support from the National Science Foundation (NSF) under Grant No. DMR-1506634.  The authors acknowledge useful discussions with N.~Ofori-Opoku, V.~Heinonen, L.~Angheluta and J.~Vi\~nals.
\end{acknowledgments}

\clearpage
\newpage

\onecolumngrid

\begin{center}
  \textbf{\large \hspace{5pt} SUPPLEMENTAL MATERIAL \\ \vspace{0.2cm} Mesoscale Defect Motion in Binary Systems: Effects of Compositional
Strain and \\ Cottrell Atmospheres}\\[.2cm]
  Marco Salvalaglio,$^{1,2,*}$, Axel Voigt,$^{1,2}$
   Zhi-Feng Huang$^{3}$, Ken R. Elder$^4$ \\[.1cm]
  {\itshape \small
  ${}^1$Institute  of Scientific Computing,  TU  Dresden,  01062  Dresden,  Germany
  \\ 
  ${}^2$Dresden Center for Computational Materials Science (DCMS), TU  Dresden,  01062  Dresden,  Germany
 \\
  ${}^3$Department of Physics and Astronomy, Wayne State University, Detroit, Michigan 48201, USA
 \\
  ${}^4$Department of Physics, Oakland University, Rochester, Michigan 48309, USA
 }
\vspace{0.5cm}
\end{center}

\setcounter{equation}{0}
\setcounter{figure}{0}
\setcounter{table}{0}
\setcounter{page}{1}
\renewcommand{\thesection}{S\arabic{section}}
\renewcommand{\theequation}{S\arabic{equation}}
\renewcommand{\thefigure}{S\arabic{figure}}
\renewcommand{\bibnumfmt}[1]{[S#1]}
\renewcommand{\citenumfont}[1]{S#1}

\section{Velocity expressions for different crystal symmetries}
\label{sec:si1}
A 2D triangular ($T$) or honeycomb lattice requires three reciprocal
vectors, $\vec{q}_1$=$\langle-\sqrt{3}/2,-1/2\rangle$, 
$\vec{q}_2$=$\langle0,1\rangle$ and $\vec{q}_3$=$ -\vec{q}_1-\vec{q}_2$, and
thus, with $(j,k)=(x,y)$,
\be
v^{\rm T}_j =\frac{4\pi \varepsilon_{jk} B_0^x}{|\vec{b}_m|^2}  \left[
2 U_{jk}b_j+(U_{jj}+3U_{kk}-4\alpha\delta\psi)b_k \right].
\label{eq:vtriSM}
\ee

A square lattice ($S$) can be represented by two modes with wave vectors
$\vec{q}_1$=$\langle1,0\rangle$, $\vec{q}_2$=$\langle0,1\rangle$,
$\vec{q}_3$=$\langle1,1\rangle$,
and $\vec{q}_4$=$\langle-1,1\rangle$.  This gives
\be
v_j^{\rm S} = \frac{16\pi\varepsilon_{jk}B_0^x}{3|\vec{b}_m|^2}\left[
    8U_{jk}b_j+(4U_{jj}+5U_{kk}-9\alpha \delta\psi)b_k
\right].
\label{eq:vsq}
\ee
A bcc crystal ($B$) requires six wave vectors of a single mode, corresponding 
to $\vec{q}_1$=$\langle1,1,0\rangle/\sqrt{2}$, 
$\vec{q}_2$=$\langle1,0,1\rangle/\sqrt{2}$, 
$\vec{q}_3$=$\langle0,1,1\rangle/\sqrt{2}$, 
$\vec{q}_4$=$\vec{q}_1-\vec{q}_2$,
$\vec{q}_5$=$\vec{q}_2-\vec{q}_3$,
and $\vec{q}_6$=$\vec{q}_3-\vec{q}_1$.  Velocity for a line defect in the $z$
direction is given by
\be
v^{\rm B}_j &=& \frac{4\pi\varepsilon_{jk} B_0^x}{|\vec{b}_m|^2}\left[
2U_{jk}b_j+ 
(U_{jj}+2U_{kk}+U_{zz}-4\alpha\delta \psi)b_k 
\right]. 
\label{eq:vbcc}
\ee
An fcc phase ($F$) can be represented by two-mode wave vectors
$\vec{q}_1$=$\langle-1,1,1\rangle/\sqrt{3}$,
$\vec{q}_2$=$\langle1,-1,1\rangle/\sqrt{3}$,
$\vec{q}_3$=$\langle1,1,-1\rangle/\sqrt{3}$,
$\vec{q}_4$=$-(\vec{q}_1+\vec{q}_2+\vec{q}_3$), 
$\vec{q}_5$=$\vec{q}_1+\vec{q}_2$, $\vec{q}_6$=$\vec{q}_2+\vec{q}_3$, and
$\vec{q}_7$=$\vec{q}_1+\vec{q}_3$, leading to
\be
v_j^{\rm F} &=& \frac{8\pi\varepsilon_{jk} B_0^x}{|9\vec{b}_m|^2}\left[
6U_{jk}b_j+(3U_{jj}+19U_{kk}+3U_{zz}-25\alpha\delta\psi)b_k \right].
\label{eq:vfcc}
\ee

\section{Model Equations and Discretization for Numerical Simulations}

We consider explicitly the partial differential equations for ${\partial \eta_n}/{\partial t}$ and ${\partial \psi}/{\partial t}$ as introduced in the main text
\begin{equation}
\begin{split}
\frac{\partial \eta_n}{\partial t}=&-|\kj|^2 \left\{ \left[
\Delta B_0 + B_2^\ell \psi^2 + B_0^x\mathcal{G}_n^2 + 3v \left(\Phi-|\eta_n |^2\right)\right]\eta_n  
 + \frac{\delta f^{\rm s}(\{\eta_n\},\{A^*_n\})}{\delta \eta_n^*} - 2 |\kj|^2 \alpha B_0^x (\psi \mathcal{G}_n \eta_n + \mathcal{G}_n \eta_n \psi) \right\}, \\
\frac{\partial \psi}{\partial t}=&\nabla^2\bigg[(w+B^\ell_2 \Phi -K\nabla^2)\psi + u\psi^3 -2 \alpha B_0^{x} \sum_n |\kj|^2(\eta_n \mathcal{G}_n^* \eta_n^* + \eta_n^* \mathcal{G}_n \eta_n)  \bigg],
\label{eq:amptimebinary}
\end{split}
\end{equation}
with $\mathcal{G}_n\equiv \nabla^2+2\im\kj \cdot \vec{\nabla}$ and $\Phi\equiv 2\sum_{n} |\eta_n|^2$. $f^{\rm s}(\{\eta_n\},\{\eta_n^*\})$ is set in agreement with the appropriate crystalline symmetry as reported in Refs.~\cite{SMElderPRE2010,SMSalvalaglioAPFC2017}, i.e. (by considering $\vec{q}_n$ as in Sec.~\ref{sec:si1} for different symmetries),
\begin{equation}
\begin{split}
    f^{\rm T} =& -2t \left(\eta_1\eta_2\eta_3 + {\rm c.c.}\right),\\
    f^{\rm S} =& -2t \left(\eta_1\eta_2\eta_3^*+\eta_1\eta_2^*\eta_4+{\rm {\rm c.c.}}\right)+3v(\eta_1\eta_1\eta_3^{*}\eta_4+
\eta_2^*\eta_2^*\eta_3\eta_4 + {\rm c.c.}),\\
    f^{\rm B} =& -2t(\eta_1^*\eta_2\eta_4+\eta_2^*\eta_3\eta_5+\eta_3^*\eta_1\eta_6 + \eta_4^*\eta_5^*\eta_6^*+{\rm c.c.})      +6v(\eta_1\eta_3^*\eta_4^*\eta_5^*+\eta_2\eta_1^*\eta_5^*\eta_6^*+\eta_3\eta_2^*\eta_6^*\eta_4^*+{\rm c.c.}),\\
    f^{\rm F} =& -2t[\eta_1^*(\eta_2^*\eta_5+\eta_3^*\eta_7+\eta_4^*\eta_6^*)+\eta_2^*(\eta_3^*\eta_6+\eta_4^*\eta_7^*)
+\eta_3^*\eta_4^*\eta_5^*+{\rm c.c.}] + 6v [\eta_1^*(\eta_2^*\eta_3^*\eta_4^*
+\eta_2 \eta_6^* \eta_7  + \eta_3\eta_5 \eta_6^*\\ 
	      &+ \eta_4\eta_5\eta_7) + \eta_2^*\eta_5(\eta_3\eta_7^*+\eta_4\eta_6) + \eta_3^*\eta_4\eta_6\eta_7+{\rm c.c.}]. 
\end{split}
\end{equation}
In our simulations the parameters are set as follows: $t=3/5$, $v=1$, $B_0^x=1$,
$\Delta B_0=0.015$, $u=4$, $w=0.088$, $K=1$, and $B_2^\ell=-1.8$.

The following implementation builds on the discretization scheme proposed in Refs.~\cite{SMSalvalaglioAPFC2017,SMPraetorius_2019}. The calculation of the time evolution of $\eta_n$ is performed by considering four differential equations for the real and imaginary parts of $\eta_n$ and $\mathcal{G}_n\eta_n$. With $\eta_n = \ra+\im \ia$ and $\mathcal{G}_n\eta_n=\rz+\im \iz$ they read
\begin{equation}
\begin{split}
\frac{\partial \ra}{\partial t}=&-|\kj|^2 \bigg\{ \Delta B_0 \ra + B_2^\ell \psi^2 \ra +B_0^{x}\nabla^2 \rz -2B_0^x\kj \cdot \vec{\nabla} \iz
+ 3v (\Phi-|\eta_n|^2) \ra + \text{Re}\left(\frac{\delta f^{\rm s}}{\delta \eta_n^*}\right) \\
&-2|\kj|^2\alpha B_0^x \left[\psi \rz + \nabla^2 (\psi \ra) - 2\kj \cdot \vec{\nabla} (\psi \ia) \right] \bigg\}, \\
\frac{\partial \ia}{\partial t}=&-|\kj|^2 \bigg\{\Delta B_0 \ia + B_2^\ell \psi^2 \ia +B_0^x\nabla^2 \iz +2B_0^x\kj \cdot \vec{\nabla} \rz  
+ 3v(\Phi-|\eta_n|^2) \ia + \text{Im}\left(\frac{\delta f^{\rm s}}{\delta \eta_n^*}\right)\\
&-2|\kj|^2\alpha B_0^x \left[\psi \iz + \nabla^2 (\psi \ia) + 2\kj \cdot \vec{\nabla} (\psi \ra)\right] \bigg\}, \\
\rz =& \nabla^2\ra - 2\kj \cdot \vec{\nabla} \ia, \\
\iz =& \nabla^2\ia + 2\kj\cdot \vec{\nabla} \ra .
\end{split}
\label{eq:eqeq}
\end{equation}
These equations should be coupled with the time evolution of $\psi$. It is also governed by a fourth order PDE and is solved by a system of two second-order PDEs as follows:
\begin{equation}
\begin{split}
\frac{\partial \psi}{\partial t}=& 2B^\ell_2 \vec{\nabla} \psi  \cdot \vec{\nabla} \Phi+ B^\ell_2 \psi \nabla^2 \Phi + 6 u \psi |\vec{\nabla} \psi|^2 + (w+B^\ell_2 \Phi +3u\psi^2)\xi - K \nabla^2 \xi 
- 4B^{x}_0\alpha \nabla^2 \left[ \sum_n |\kj|^2 (\ra\rz +\ia\iz)\right],\\
\xi=&\nabla^2 \psi .
\end{split}
\label{eq:eqeq2}
\end{equation}

Let us consider the time discretization $t_\iota$ with $\iota \in \mathbb{N}$ such as $0=t_0<t_1< ...$ and the time step $\tau_\iota=t_{\iota+1}-t_{\iota}$. The semi-implicit integration scheme for $\partial \eta_n/\partial t$ in the matrix form $\mathcal{L}_{\eta} \cdot \mathcal{X}_{\eta}=\mathcal{R}_{\eta}$ reads
\begin{equation}
\mathcal{L}_{\eta}=
\begin{bmatrix}
-\nabla^2 & \mathcal{A} &1 &0 \\[0.65em]
-\mathcal{A} &-\nabla^2 &0 &1 \\[0.65em]
G_1(\{\eta_n^{(\iota)}\},\psi^{(\iota)}) & 0 & \mathcal{K} (\nabla^2 -2|\kj|^2\alpha \psi^{(\iota)}) & -\mathcal{K} \mathcal{A}  \\[0.65em]
0 & G_2(\{\eta_n^{(\iota)}\},\psi^{(\iota)}) & \mathcal{K} \mathcal{A} & \mathcal{K} (\nabla^2 -2|\kj|^2\alpha \psi^{(\iota)}) 
\end{bmatrix}, 
\label{eq:integrationscheme1a}
\end{equation}
\begin{equation}
 \mathcal{X}_{\eta}=
\begin{bmatrix}
\ra^{(\iota+1)} \\[0.75em]
\ia^{(\iota+1)} \\[0.75em] 
\rz^{(\iota+1)}  \\[0.75em]
\iz^{(\iota+1)}, 
\end{bmatrix}, \qquad
\mathcal{R}_{\eta}= 
\begin{bmatrix}
0\\[0.75em]
0\\[0.75em] 
H_1(\{\eta_n^{(\iota)},\psi^{(\iota)}\})\\[0.75em]
H_2(\{\eta_n^{(\iota)},\psi^{(\iota)}\}) 
\end{bmatrix},
\label{eq:integrationscheme1b}
\end{equation}

where $\mathcal{A}=2\kj \cdot \vec{\nabla}$ and $\mathcal{K}=|\kj|^2B_0^x$, while the functions evaluated explicitly at time $t_n$ are given by
\begin{equation}
\begin{split}
G_1(\{\eta_n\},\psi)=&\dfrac{1}{\tau_\iota}+|\kj|^2\Delta B_0+3v|\kj|^2\left(\Phi+\ra^2-\ia^2\right)+|\kj|^2 B_2^\ell \psi^2,\\
G_2(\{\eta_n\},\psi)=&\dfrac{1}{\tau_\iota}+|\kj|^2\Delta B_0+3v|\kj|^2\left(\Phi+\ia^2-\ra^2\right)+|\kj|^2 B_2^\ell \psi^2, \\
H_1(\{\eta_n\},\psi)=&\left[ \dfrac{1}{\tau_\iota}+6|\kj|^2v \ra^2 \right]\ra -|\kj|^2\text{Re}\left(\frac{\delta f^{s}}{\delta \eta_n^*}\right) 
+2|\kj|^4\alpha B_0^x \left[\nabla^2 (\psi \ra) - 2\kj \cdot \vec{\nabla} (\psi \ia) \right], \\
H_2(\{\eta_n\},\psi)=&\left[\dfrac{1}{\tau_\iota} +6|\kj|^2v \ia^2 \right]\ia -|\kj|^2\text{Im}\left(\frac{\delta f^{s}}{\delta \eta_n^*}\right) 
+2|\kj|^4\alpha B_0^x \left[\nabla^2 (\psi \ia) + 2\kj \cdot \vec{\nabla} (\psi \ra) \right]. \\
\end{split}
\label{eq:system1}
\end{equation}
The functions in Eq.~\eqref{eq:system1} account for the right- and left-hand side terms resulting from the linearization of $-3v \left( \Phi-|\eta_n|^2 \right)\ra$ and $-3v \left( \Phi-|\eta_n|^2 \right)\ia$ terms in Eq.~\eqref{eq:eqeq} with respect to $\ra^{(\iota+1)}$ and $\ia^{(\iota+1)}$ around $\ra^{(\iota)}$ and $\ia^{(\iota)}$, respectively. In order to compute the evolution of the amplitudes, the system defined by Eqs. \eqref{eq:integrationscheme1a} and \eqref{eq:integrationscheme1b} needs to be solved for each $\eta_n$, which involves the coupling between different amplitudes through $f^{\rm s}$. 

The semi-implicit integration scheme used here to compute $\partial \psi/\partial t$ (in the matrix form $\mathcal{L}_{\psi} \cdot \mathcal{X}_{\psi}=\mathcal{R}_{\psi}$) reads
\begin{equation}
\mathcal{L}_{\psi}=
\begin{bmatrix}
\nabla^2 & -1 \\[0.75em]
[Q(\{\eta_n^{(\iota)}\},\psi^{(\iota)}) + \vec{M}(\{\eta_n^{(\iota)}\})  \cdot \vec{\nabla}]  & [O(\{\eta_n^{(\iota)}\},\psi^{(\iota)}) + K\nabla^2] \\
\end{bmatrix}, 
\label{eq:integrationscheme2a}
\end{equation}
\begin{equation}
 \mathcal{X}_{\psi}=
\begin{bmatrix}
\psi^{(\iota+1)} \\[0.75em]
\xi^{(\iota+1)} 
\end{bmatrix}, \qquad
\mathcal{R}_{\psi}= 
\begin{bmatrix}
0\\[0.75em]
P(\{\eta_n^{(\iota)}\},\psi^{(\iota)}) 
\end{bmatrix},
\label{eq:integrationscheme2b}
\end{equation}
where
\begin{equation}
\begin{split}
Q(\{\eta_n\},\psi)=&\frac{1}{\tau_\iota}-B_2^\ell \nabla^2 \Phi - 6u|\vec{\nabla} \psi|^2,\\
\vec{M}(\{\eta_n\})=&-2B_2^\ell \vec{\nabla} \Phi,\\
O(\{\eta_n\},\psi)=&-(w+B^\ell_2 \Phi+3u\psi^2),\\
P(\{\eta_n\},\psi)=&\frac{\psi}{\tau_\iota} - 4B^{x}_0\alpha \nabla^2 \left[ \sum_n |\kj|^2 (\ra\rz+\ia\iz)\right].
\end{split}
\label{eq:system}
\end{equation}

The integration schemes reported above have been implemented in the Finite Element Method framework available within the AMDiS toolbox \cite{SMVey2007,SMWitkowski2015}. All the second-order, explicit terms entering the systems at the right hand side (i.e. in $\mathcal{R}_{\eta,\psi}$) are evaluated by first computing gradients on the variables and then calculating the first-order terms in the weak form within the considered FEM approach.

\section{Initial Conditions for Amplitudes}

The initial conditions for amplitudes are set through the following equation,
\begin{equation}
    \eta_n=\phi_n e^{-\im(\vec{q}_n^{\,\rm eq}-\vec{q}_n)\cdot \vec{r}}e^{-\im\vec{q}_n^{\,\rm eq}\cdot\vec{u}},
    \label{eq:ini}
\end{equation}
with $\vec{u}$ the targeted lattice distortion to be encoded, $\vec{q}_n^{\,\rm eq}=\vec{q}_n\sqrt{1-2\alpha\psi}$ \cite{SMHuang10}, and $\phi_n$ the real amplitudes describing a relaxed bulk crystal with values depending on lattice symmetry and parameters entering the energy \cite{SMElderPRE2010,SMSalvalaglioAPFC2017}. Note that if $\vec{q}_n^{\,\rm eq}=\vec{q}_n$, e.g., for $\alpha=0$ or $\bar{\psi}=0$, Eq.~\eqref{eq:ini} reduces to the previous expression for amplitudes of a deformed lattice $\eta_n=\phi_n e^{-\im\vec{q}_n\cdot\vec{u}}$ \cite{SMElderPRE2010,SMSalvalaglioAPFC2017}.

The initial condition exploited for Figs.~1 and 2(a) in the main text is set by initializing the amplitudes through $\vec{u}=(u_x,u_y)=(\text{Sign}(L_y/2-|y|)a_{\rm tri}/L_x,0)$ with $L_x = L_y \sim 500a_{\rm tri}$ and $100a_{\rm tri}$, respectively, for a 2D triangular structure \cite{SMSalvalaglioAPFC2017}. Periodic Boundary Conditions can be applied by considering a simulation domain with size matching amplitude periodicity.

For the 3D configurations leading to the results reported in Figure 2(b), we set $\hat{x}= [100]$, $\vec{u}=(u_x,u_y,u_z)=(\text{Sign}(L_y/2-|y|)a_{100}^{B}/L_x,0,0)$, and $a_{100}^{B}=2\pi\sqrt{2}$ for the bcc case, and $\hat{x}= [110]$, $\vec{u}=(u_x,u_y,u_z)=(\text{Sign}(L_y/2-|y|)a_{110}^{F}/L_x,0,0)$, and $a_{110}^{F}=\pi/\sqrt{6}$ for the fcc case. Here $L_x = L_y = L_z \sim 50\pi$.

The configurations G and C used for the results illustrated in Figure 3 are obtained by explicitly considering the displacement field generated by two dislocations at the corresponding positions. The components of the displacement field for an edge dislocation with Burgers vector pointing along the $x$ direction are given by \cite{SManderson2017}:
\begin{equation}
 \begin{split}
u_x(\bar{r},b)=&\frac{b}{2\pi} \left[\tan^{-1} \left(\frac{\bar{y}}{\bar{x}}\right)+\frac{\bar{x}\bar{y}}{2(1-\nu)(\bar{x}^2+\bar{y}^2)} \right], \\
u_y({\bar{r}},b)=&-\frac{b}{2\pi} \left[\frac{(1-2\nu)}{4(1-\nu)}\log(\bar{x}^2+\bar{y}^2) + \frac{\bar{x}^2-\bar{y}^2}{4(1-\nu)(\bar{x}^2+\bar{y}^2)} \right],  
\label{eq:u_dislo}
\end{split}
\end{equation}
where $b=a_{\rm tri}=4 \pi / \sqrt{3}$ for the 2D triangular crystal examined, and $\bar{r}=(\bar{x},\bar{y})=(x-x_0,y-y_0)$ corresponds to the coordinates shifted with respect to the dislocation core. Singularities at $(\bar{x},\bar{y})=(0,0)$ are removed as in Ref.~\cite{SMcai2006non}. Configuration G can be obtained by setting $\vec{u}_G=\vec{u}(x-x_0,y,-b)+\vec{u}(x+x_0,y,b)$, while for configuration C,
$\vec{u}_C=\vec{u}(x,y-y_0,b)+\vec{u}(x,y+y_0,-b)$.

The initial condition for the simulations illustrated in Figure 4 for bcc grain shrinkage is obtained by setting the displacement field as $\vec{u}= [\mathcal{R}(\vartheta)-1]\vec{r}$, where $\mathcal{R}(\vartheta)$ is the counter-clockwise rotation matrix, and $\vartheta=\vartheta(\vec{r}) = \vartheta_g\Theta(R_0-|\vec{r}|)$ with $\vartheta_g$ the grain misorientation angle and $R_0$ the initial grain radius.

\section{Stress and strain}

Expressions for the stress and strain as a function of amplitudes $\eta_n$ can be derived for a pure crystal as described in Refs.~\cite{SMSalvalaglioNPJ2019,SMSalvalaglioJMPS2020}. However, in the presence of a solute, namely for binary systems with $\alpha \neq 0$, additional contributions must be included. Consider the last term of Eq.~(3) in the main text, which couples the solute concentration field with amplitudes $\eta_n$ and their spatial variation, i.e.,
\be
f_{\psi} = -2B_0^x\alpha\sum_n q_n^2
\left(\eta_n {\cal G}_n^*\eta_n^*+{\rm c.c.}\right)\psi.
\ee
Using $\mathcal{G}_n=\nabla^2+2\im \vec{q}_n\cdot \vec{\nabla}$ and $\eta_n = \phi_n e^{-\im\vec{q}_n\cdot\vec{u}}$ we get 
\begin{equation}
\begin{split}
f_{\psi} 
&\approx 8B_0^x\alpha\psi \sum_n |\phi_n|^2 q_n^2  {q}^{n}_{i} {q}^{n}_{j} \partial_j u_i,
\end{split}
\end{equation}
where higher order terms [$\sim (\partial_j u_i)^2$] have been neglected. By exploiting the definition of strain, $U_{ij}=\frac{1}{2}\left( \partial_i{u_j} +\partial_j{u_i} \right)$, 
we can obtain the compositional stress via $\sigma_{ij}^{\psi}=\partial f_{\psi} / \partial U_{ij}$. The total stress is then given by $\sigma_{ij}=\sigma_{ij}^{\rm \eta}+\sigma_{ij}^{\psi}$, with $\sigma_{ij}^{\rm \eta}$ the stress components related to lattice distortion (see Ref.~\cite{SMSalvalaglioJMPS2020}). Strain components can generally be computed as $U_{ij}=S_{ijkl}\sigma_{kl}$ with $S_{ijkl}$ the compliance tensor. For isotropic materials and 2D hexagonal crystals the following equation is used
\begin{equation}
    U_{ij}=\frac{1}{2\mu} \sigma_{ij}-\frac{\lambda}{2\mu(d\lambda+2\mu)}\delta_{ij}\sigma_{kk},
    \label{eq:Usigma}
\end{equation}
with $\lambda$ and $\mu$ the Lam\'e coefficients and $d$ the system dimensionality.

For the case of triangular symmetry illustrated in Figure 1 (see the corresponding definition of $\vec{q}_{n}$ in Sect.~\ref{sec:si1}) we obtain
\be
f_{\psi} &=& 6\alpha B_0^x\left(|\phi_1|^2+|\phi_3|^2\right)\psi U_{xx}
+ 2\alpha B_0^x\left(|\phi_1|^2+4|\phi_2|^2+|\phi_3|^2\right)\psi U_{yy}
 + 4\sqrt{3}\alpha B_0^x\left(|\phi_1|^2-|\phi_3|^2\right)\psi U_{xy},
\ee
leading to the following results of $\sigma_{ij}^{\psi}$
\be
\sigma_{xx}^{\psi}&=&  6\alpha B_0^x\left(|\phi_1|^2+|\phi_3|^2\right)\psi,  \nline
\sigma_{yy}^{\psi} &=& 2\alpha B_0^x\left(|\phi_1|^2+4|\phi_2|^2+|\phi_3|^2\right)\psi,  \nline
\sigma_{xy}^{\psi} &=& 4\sqrt{3}\alpha B_0^x\left(|\phi_1|^2-|\phi_3|^2\right)\psi.
\ee
$U_{ij}$ can be computed by Eq.~\eqref{eq:Usigma} with $\lambda=\mu=3\phi^2B_0^x$ (where $\phi$ is the real, constant amplitude describing a relaxed crystal with $\vec{u}=0$). Notice that for the triangular symmetry, and generally for symmetries described by $\vec{q}_n$ wave vectors of equal length, $\phi_n=\phi$ in bulk, while $\phi_n$ vary differently for different component $n$ around defects and interfaces.


\begin{thebibliography}{57}%
\makeatletter
\providecommand \@ifxundefined [1]{%
 \@ifx{#1\undefined}
}%
\providecommand \@ifnum [1]{%
 \ifnum #1\expandafter \@firstoftwo
 \else \expandafter \@secondoftwo
 \fi
}%
\providecommand \@ifx [1]{%
 \ifx #1\expandafter \@firstoftwo
 \else \expandafter \@secondoftwo
 \fi
}%
\providecommand \natexlab [1]{#1}%
\providecommand \enquote  [1]{``#1''}%
\providecommand \bibnamefont  [1]{#1}%
\providecommand \bibfnamefont [1]{#1}%
\providecommand \citenamefont [1]{#1}%
\providecommand \href@noop [0]{\@secondoftwo}%
\providecommand \href [0]{\begingroup \@sanitize@url \@href}%
\providecommand \@href[1]{\@@startlink{#1}\@@href}%
\providecommand \@@href[1]{\endgroup#1\@@endlink}%
\providecommand \@sanitize@url [0]{\catcode `\\12\catcode `\$12\catcode
  `\&12\catcode `\#12\catcode `\^12\catcode `\_12\catcode `\%12\relax}%
\providecommand \@@startlink[1]{}%
\providecommand \@@endlink[0]{}%
\providecommand \url  [0]{\begingroup\@sanitize@url \@url }%
\providecommand \@url [1]{\endgroup\@href {#1}{\urlprefix }}%
\providecommand \urlprefix  [0]{URL }%
\providecommand \Eprint [0]{\href }%
\providecommand \doibase [0]{https://doi.org/}%
\providecommand \selectlanguage [0]{\@gobble}%
\providecommand \bibinfo  [0]{\@secondoftwo}%
\providecommand \bibfield  [0]{\@secondoftwo}%
\providecommand \translation [1]{[#1]}%
\providecommand \BibitemOpen [0]{}%
\providecommand \bibitemStop [0]{}%
\providecommand \bibitemNoStop [0]{.\EOS\space}%
\providecommand \EOS [0]{\spacefactor3000\relax}%
\providecommand \BibitemShut  [1]{\csname bibitem#1\endcsname}%
\let\auto@bib@innerbib\@empty
\bibitem [{\citenamefont {Herzer}(2013)}]{Herzer2013}%
  \BibitemOpen
  \bibfield  {author} {\bibinfo {author} {\bibfnamefont {G.}~\bibnamefont
  {Herzer}},\ }\bibfield  {title} {\bibinfo {title} {Modern soft magentic:
  Amorphous and nanocrystalline materials},\ }\href
  {https://doi.org/10.1016/j.actamat.2012.10.040} {\bibfield  {journal}
  {\bibinfo  {journal} {Acta. Mater.}\ }\textbf {\bibinfo {volume} {61}},\
  \bibinfo {pages} {718} (\bibinfo {year} {2013})}\BibitemShut {NoStop}%
\bibitem [{\citenamefont {Xue}\ \emph {et~al.}(2008)\citenamefont {Xue},
  \citenamefont {Chai}, \citenamefont {Li},\ and\ \citenamefont
  {Fan}}]{Xue2008}%
  \BibitemOpen
  \bibfield  {author} {\bibinfo {author} {\bibfnamefont {D.}~\bibnamefont
  {Xue}}, \bibinfo {author} {\bibfnamefont {G.}~\bibnamefont {Chai}}, \bibinfo
  {author} {\bibfnamefont {X.}~\bibnamefont {Li}},\ and\ \bibinfo {author}
  {\bibfnamefont {X.}~\bibnamefont {Fan}},\ }\bibfield  {title} {\bibinfo
  {title} {Effects of grain size distributions on coercivity and permeability
  of ferromagnets},\ }\href {https://doi.org/10.1016/j.jmmm.2008.01.004}
  {\bibfield  {journal} {\bibinfo  {journal} {J. Mag. Mag. Mat.}\ }\textbf
  {\bibinfo {volume} {320}},\ \bibinfo {pages} {1541} (\bibinfo {year}
  {2008})}\BibitemShut {NoStop}%
\bibitem [{\citenamefont {Yip}(1998)}]{Yip1998}%
  \BibitemOpen
  \bibfield  {author} {\bibinfo {author} {\bibfnamefont {S.}~\bibnamefont
  {Yip}},\ }\bibfield  {title} {\bibinfo {title} {The strongest size},\ }\href
  {https://doi.org/10.1038/35254} {\bibfield  {journal} {\bibinfo  {journal}
  {Nature}\ }\textbf {\bibinfo {volume} {391}},\ \bibinfo {pages} {532}
  (\bibinfo {year} {1998})}\BibitemShut {NoStop}%
\bibitem [{\citenamefont {Petch}(1953)}]{Petch1953}%
  \BibitemOpen
  \bibfield  {author} {\bibinfo {author} {\bibfnamefont {N.~J.}\ \bibnamefont
  {Petch}},\ }\bibfield  {title} {\bibinfo {title} {The cleavage strength of
  polycrystals},\ }\href@noop {} {\bibfield  {journal} {\bibinfo  {journal} {J.
  Iron Steel Inst., London}\ }\textbf {\bibinfo {volume} {174}},\ \bibinfo
  {pages} {25} (\bibinfo {year} {1953})}\BibitemShut {NoStop}%
\bibitem [{\citenamefont {Fan}\ \emph {et~al.}(2017)\citenamefont {Fan},
  \citenamefont {Hirvonen}, \citenamefont {Pereira}, \citenamefont {Ervasti},
  \citenamefont {Elder}, \citenamefont {Donadio}, \citenamefont {Harju},\ and\
  \citenamefont {Ala-Nissila}}]{Fan2017}%
  \BibitemOpen
  \bibfield  {author} {\bibinfo {author} {\bibfnamefont {Z.}~\bibnamefont
  {Fan}}, \bibinfo {author} {\bibfnamefont {P.}~\bibnamefont {Hirvonen}},
  \bibinfo {author} {\bibfnamefont {L.~F.~C.}\ \bibnamefont {Pereira}},
  \bibinfo {author} {\bibfnamefont {M.~M.}\ \bibnamefont {Ervasti}}, \bibinfo
  {author} {\bibfnamefont {K.~R.}\ \bibnamefont {Elder}}, \bibinfo {author}
  {\bibfnamefont {D.}~\bibnamefont {Donadio}}, \bibinfo {author} {\bibfnamefont
  {A.}~\bibnamefont {Harju}},\ and\ \bibinfo {author} {\bibfnamefont
  {T.}~\bibnamefont {Ala-Nissila}},\ }\bibfield  {title} {\bibinfo {title}
  {Bimodal grain-size scaling of thermal transport in polycrystalline graphene
  from large-scale molecular dynamics simulations},\ }\href
  {https://doi.org/10.1021/acs.nanolett.7b01742} {\bibfield  {journal}
  {\bibinfo  {journal} {Nano Lett.}\ }\textbf {\bibinfo {volume} {17}},\
  \bibinfo {pages} {5919} (\bibinfo {year} {2017})}\BibitemShut {NoStop}%
\bibitem [{\citenamefont {Cottrell}\ \emph {et~al.}(1949)\citenamefont
  {Cottrell}, \citenamefont {Jaswon},\ and\ \citenamefont
  {Mott}}]{Cottrell1949}%
  \BibitemOpen
  \bibfield  {author} {\bibinfo {author} {\bibfnamefont {A.~H.}\ \bibnamefont
  {Cottrell}}, \bibinfo {author} {\bibfnamefont {M.~A.}\ \bibnamefont
  {Jaswon}},\ and\ \bibinfo {author} {\bibfnamefont {N.~F.}\ \bibnamefont
  {Mott}},\ }\bibfield  {title} {\bibinfo {title} {Distribution of solute atoms
  round a slow dislocation},\ }\href {https://doi.org/10.1098/rspa.1949.0128}
  {\bibfield  {journal} {\bibinfo  {journal} {Proc. R. Soc. Lond. A}\ }\textbf
  {\bibinfo {volume} {199}},\ \bibinfo {pages} {104} (\bibinfo {year}
  {1949})}\BibitemShut {NoStop}%
\bibitem [{\citenamefont {Cottrell}\ and\ \citenamefont
  {Bilby}(1949)}]{Cottrell1949_2}%
  \BibitemOpen
  \bibfield  {author} {\bibinfo {author} {\bibfnamefont {A.~H.}\ \bibnamefont
  {Cottrell}}\ and\ \bibinfo {author} {\bibfnamefont {B.~A.}\ \bibnamefont
  {Bilby}},\ }\bibfield  {title} {\bibinfo {title} {Dislocation theory of
  yielding and strain ageing of iron},\ }\href
  {https://doi.org/10.1088/0370-1298/62/1/308} {\bibfield  {journal} {\bibinfo
  {journal} {Proc. R. Soc. Lond. A}\ }\textbf {\bibinfo {volume} {62}},\
  \bibinfo {pages} {49} (\bibinfo {year} {1949})}\BibitemShut {NoStop}%
\bibitem [{\citenamefont {Cottrell}(1953)}]{CottrellBOOK}%
  \BibitemOpen
  \bibfield  {author} {\bibinfo {author} {\bibfnamefont {A.~H.}\ \bibnamefont
  {Cottrell}},\ }\href {https://cds.cern.ch/record/503960} {\emph {\bibinfo
  {title} {{Dislocations and plastic flow in crystals}}}},\ Internat. Ser.
  Mono. Phys.\ (\bibinfo  {publisher} {Clarendon Press},\ \bibinfo {address}
  {Oxford},\ \bibinfo {year} {1953})\BibitemShut {NoStop}%
\bibitem [{\citenamefont {L{\"u}cke}\ and\ \citenamefont
  {Detert}(1957)}]{LUCKE1957}%
  \BibitemOpen
  \bibfield  {author} {\bibinfo {author} {\bibfnamefont {K.}~\bibnamefont
  {L{\"u}cke}}\ and\ \bibinfo {author} {\bibfnamefont {K.}~\bibnamefont
  {Detert}},\ }\bibfield  {title} {\bibinfo {title} {A quantitative theory of
  grain-boundary motion and recrystallization in metals in the presence of
  impurities},\ }\href
  {https://doi.org/https://doi.org/10.1016/0001-6160(57)90109-8} {\bibfield
  {journal} {\bibinfo  {journal} {Acta Metall.}\ }\textbf {\bibinfo {volume}
  {5}},\ \bibinfo {pages} {628 } (\bibinfo {year} {1957})}\BibitemShut
  {NoStop}%
\bibitem [{\citenamefont {Cahn}(1962)}]{CAHN1962}%
  \BibitemOpen
  \bibfield  {author} {\bibinfo {author} {\bibfnamefont {J.~W.}\ \bibnamefont
  {Cahn}},\ }\bibfield  {title} {\bibinfo {title} {The impurity-drag effect in
  grain boundary motion},\ }\href
  {https://doi.org/https://doi.org/10.1016/0001-6160(62)90092-5} {\bibfield
  {journal} {\bibinfo  {journal} {Acta Metall.}\ }\textbf {\bibinfo {volume}
  {10}},\ \bibinfo {pages} {789 } (\bibinfo {year} {1962})}\BibitemShut
  {NoStop}%
\bibitem [{\citenamefont {L{\"u}cke}\ and\ \citenamefont
  {St{\"u}we}(1971)}]{LUCKE1971}%
  \BibitemOpen
  \bibfield  {author} {\bibinfo {author} {\bibfnamefont {K.}~\bibnamefont
  {L{\"u}cke}}\ and\ \bibinfo {author} {\bibfnamefont {H.}~\bibnamefont
  {St{\"u}we}},\ }\bibfield  {title} {\bibinfo {title} {On the theory of
  impurity controlled grain boundary motion},\ }\href
  {https://doi.org/https://doi.org/10.1016/0001-6160(71)90041-1} {\bibfield
  {journal} {\bibinfo  {journal} {Acta Metall.}\ }\textbf {\bibinfo {volume}
  {19}},\ \bibinfo {pages} {1087 } (\bibinfo {year} {1971})}\BibitemShut
  {NoStop}%
\bibitem [{\citenamefont {Hillert}\ and\ \citenamefont
  {Sundman}(1976)}]{HILLERT1976}%
  \BibitemOpen
  \bibfield  {author} {\bibinfo {author} {\bibfnamefont {M.}~\bibnamefont
  {Hillert}}\ and\ \bibinfo {author} {\bibfnamefont {B.}~\bibnamefont
  {Sundman}},\ }\bibfield  {title} {\bibinfo {title} {A treatment of the solute
  drag on moving grain boundaries and phase interfaces in binary alloys},\
  }\href {https://doi.org/https://doi.org/10.1016/0001-6160(76)90108-5}
  {\bibfield  {journal} {\bibinfo  {journal} {Acta Metall.}\ }\textbf {\bibinfo
  {volume} {24}},\ \bibinfo {pages} {731 } (\bibinfo {year}
  {1976})}\BibitemShut {NoStop}%
\bibitem [{\citenamefont {Hillert}(2004)}]{HILLERT2004}%
  \BibitemOpen
  \bibfield  {author} {\bibinfo {author} {\bibfnamefont {M.}~\bibnamefont
  {Hillert}},\ }\bibfield  {title} {\bibinfo {title} {Solute drag in grain
  boundary migration and phase transformations},\ }\href
  {https://doi.org/https://doi.org/10.1016/j.actamat.2004.07.032} {\bibfield
  {journal} {\bibinfo  {journal} {Acta Mater.}\ }\textbf {\bibinfo {volume}
  {52}},\ \bibinfo {pages} {5289 } (\bibinfo {year} {2004})}\BibitemShut
  {NoStop}%
\bibitem [{\citenamefont {Cahn}(2013)}]{Cahn2013}%
  \BibitemOpen
  \bibfield  {author} {\bibinfo {author} {\bibfnamefont {J.~W.}\ \bibnamefont
  {Cahn}},\ }\bibfield  {title} {\bibinfo {title} {Thermodynamic aspects of
  {Cottrell} atmospheres},\ }\href
  {https://doi.org/10.1080/14786435.2013.793853} {\bibfield  {journal}
  {\bibinfo  {journal} {Philos. Mag.}\ }\textbf {\bibinfo {volume} {93}},\
  \bibinfo {pages} {3741} (\bibinfo {year} {2013})}\BibitemShut {NoStop}%
\bibitem [{\citenamefont {Hirth}(2014{\natexlab{a}})}]{Hirth2014}%
  \BibitemOpen
  \bibfield  {author} {\bibinfo {author} {\bibfnamefont {J.~P.}\ \bibnamefont
  {Hirth}},\ }\bibfield  {title} {\bibinfo {title} {On definitions and
  assumptions in the dislocation theory for solid solutions},\ }\href
  {https://doi.org/0.1080/14786435.2014.951707} {\bibfield  {journal} {\bibinfo
   {journal} {Philos. Mag.}\ }\textbf {\bibinfo {volume} {94}},\ \bibinfo
  {pages} {3162} (\bibinfo {year} {2014}{\natexlab{a}})}\BibitemShut {NoStop}%
\bibitem [{\citenamefont {Cahn}(2014)}]{Cahn2014}%
  \BibitemOpen
  \bibfield  {author} {\bibinfo {author} {\bibfnamefont {J.~W.}\ \bibnamefont
  {Cahn}},\ }\bibfield  {title} {\bibinfo {title} {Reprise: partial chemical
  strain dislocations and their role in pinning dislocations to their
  atmospheres},\ }\href {https://doi.org/10.1080/14786435.2014.951711}
  {\bibfield  {journal} {\bibinfo  {journal} {Philos. Mag.}\ }\textbf {\bibinfo
  {volume} {94}},\ \bibinfo {pages} {3170} (\bibinfo {year}
  {2014})}\BibitemShut {NoStop}%
\bibitem [{\citenamefont {Hirth}(2014{\natexlab{b}})}]{Hirth2014b}%
  \BibitemOpen
  \bibfield  {author} {\bibinfo {author} {\bibfnamefont {J.~P.}\ \bibnamefont
  {Hirth}},\ }\bibfield  {title} {\bibinfo {title} {Response to comments},\
  }\href {https://doi.org/10.1080/14786435.2014.952255} {\bibfield  {journal}
  {\bibinfo  {journal} {Philos. Mag.}\ }\textbf {\bibinfo {volume} {94}},\
  \bibinfo {pages} {3177} (\bibinfo {year} {2014}{\natexlab{b}})}\BibitemShut
  {NoStop}%
\bibitem [{\citenamefont {Mishin}\ and\ \citenamefont
  {Cahn}(2016)}]{MISHIN2016}%
  \BibitemOpen
  \bibfield  {author} {\bibinfo {author} {\bibfnamefont {Y.}~\bibnamefont
  {Mishin}}\ and\ \bibinfo {author} {\bibfnamefont {J.~W.}\ \bibnamefont
  {Cahn}},\ }\bibfield  {title} {\bibinfo {title} {Thermodynamics of {Cottrell}
  atmospheres tested by atomistic simulations},\ }\href
  {https://doi.org/https://doi.org/10.1016/j.actamat.2016.07.013} {\bibfield
  {journal} {\bibinfo  {journal} {Acta Mater.}\ }\textbf {\bibinfo {volume}
  {117}},\ \bibinfo {pages} {197 } (\bibinfo {year} {2016})}\BibitemShut
  {NoStop}%
\bibitem [{\citenamefont {Hirth}\ \emph {et~al.}(2017)\citenamefont {Hirth},
  \citenamefont {Barnett},\ and\ \citenamefont {Hoagland}}]{HIRTH2017}%
  \BibitemOpen
  \bibfield  {author} {\bibinfo {author} {\bibfnamefont {J.~P.}\ \bibnamefont
  {Hirth}}, \bibinfo {author} {\bibfnamefont {D.~M.}\ \bibnamefont {Barnett}},\
  and\ \bibinfo {author} {\bibfnamefont {R.~G.}\ \bibnamefont {Hoagland}},\
  }\bibfield  {title} {\bibinfo {title} {Solute atmospheres at dislocations},\
  }\href {https://doi.org/https://doi.org/10.1016/j.actamat.2017.03.014}
  {\bibfield  {journal} {\bibinfo  {journal} {Acta Mater.}\ }\textbf {\bibinfo
  {volume} {131}},\ \bibinfo {pages} {574 } (\bibinfo {year}
  {2017})}\BibitemShut {NoStop}%
\bibitem [{\citenamefont {Yoshinaga}\ and\ \citenamefont
  {Morozumi}(1971)}]{Yoshinaga1971}%
  \BibitemOpen
  \bibfield  {author} {\bibinfo {author} {\bibfnamefont {H.}~\bibnamefont
  {Yoshinaga}}\ and\ \bibinfo {author} {\bibfnamefont {S.}~\bibnamefont
  {Morozumi}},\ }\bibfield  {title} {\bibinfo {title} {The solute atmosphere
  round a moving dislocation and its dragging stress},\ }\href
  {https://doi.org/10.1080/14786437108217008} {\bibfield  {journal} {\bibinfo
  {journal} {Philos. Mag. A}\ }\textbf {\bibinfo {volume} {23}},\ \bibinfo
  {pages} {1367} (\bibinfo {year} {1971})}\BibitemShut {NoStop}%
\bibitem [{\citenamefont {Takeuchi}\ and\ \citenamefont
  {Argon}(1979)}]{Takeuchi1979}%
  \BibitemOpen
  \bibfield  {author} {\bibinfo {author} {\bibfnamefont {S.}~\bibnamefont
  {Takeuchi}}\ and\ \bibinfo {author} {\bibfnamefont {A.~S.}\ \bibnamefont
  {Argon}},\ }\bibfield  {title} {\bibinfo {title} {Glide and climb resistance
  to the motion of an edge dislocation due to dragging a {Cottrell}
  atmosphere},\ }\href {https://doi.org/10.1080/01418617908234833} {\bibfield
  {journal} {\bibinfo  {journal} {Philos. Mag. A}\ }\textbf {\bibinfo {volume}
  {40}},\ \bibinfo {pages} {65} (\bibinfo {year} {1979})}\BibitemShut {NoStop}%
\bibitem [{\citenamefont {Zhang}\ and\ \citenamefont
  {Curtin}(2008)}]{Zhang_2008}%
  \BibitemOpen
  \bibfield  {author} {\bibinfo {author} {\bibfnamefont {F.}~\bibnamefont
  {Zhang}}\ and\ \bibinfo {author} {\bibfnamefont {W.~A.}\ \bibnamefont
  {Curtin}},\ }\bibfield  {title} {\bibinfo {title} {Atomistically informed
  solute drag in {Al{\textendash}Mg}},\ }\href
  {https://doi.org/10.1088/0965-0393/16/5/055006} {\bibfield  {journal}
  {\bibinfo  {journal} {Model. Simul. Mater. Sci. Eng.}\ }\textbf {\bibinfo
  {volume} {16}},\ \bibinfo {pages} {055006} (\bibinfo {year}
  {2008})}\BibitemShut {NoStop}%
\bibitem [{\citenamefont {Sills}\ and\ \citenamefont {Cai}(2016)}]{Sills2016}%
  \BibitemOpen
  \bibfield  {author} {\bibinfo {author} {\bibfnamefont {R.~B.}\ \bibnamefont
  {Sills}}\ and\ \bibinfo {author} {\bibfnamefont {W.}~\bibnamefont {Cai}},\
  }\bibfield  {title} {\bibinfo {title} {Solute drag on perfect and extended
  dislocations},\ }\href {https://doi.org/10.1080/14786435.2016.1142677}
  {\bibfield  {journal} {\bibinfo  {journal} {Philos. Mag.}\ }\textbf {\bibinfo
  {volume} {96}},\ \bibinfo {pages} {895} (\bibinfo {year} {2016})}\BibitemShut
  {NoStop}%
\bibitem [{\citenamefont {Gu}\ and\ \citenamefont {El-Awady}(2020)}]{Gu2020}%
  \BibitemOpen
  \bibfield  {author} {\bibinfo {author} {\bibfnamefont {Y.}~\bibnamefont
  {Gu}}\ and\ \bibinfo {author} {\bibfnamefont {J.~A.}\ \bibnamefont
  {El-Awady}},\ }\bibfield  {title} {\bibinfo {title} {{Theoretical framework
  for predicting solute concentrations and solute-induced stresses in finite
  volumes with arbitrary elastic fields}},\ }\href
  {https://doi.org/10.1186/s41313-020-00020-2} {\bibfield  {journal} {\bibinfo
  {journal} {Mater. Theory}\ }\textbf {\bibinfo {volume} {4}},\ \bibinfo
  {pages} {1} (\bibinfo {year} {2020})}\BibitemShut {NoStop}%
\bibitem [{\citenamefont {Mishin}(2019)}]{MISHIN2019}%
  \BibitemOpen
  \bibfield  {author} {\bibinfo {author} {\bibfnamefont {Y.}~\bibnamefont
  {Mishin}},\ }\bibfield  {title} {\bibinfo {title} {Solute drag and dynamic
  phase transformations in moving grain boundaries},\ }\href
  {https://doi.org/https://doi.org/10.1016/j.actamat.2019.08.046} {\bibfield
  {journal} {\bibinfo  {journal} {Acta Mater.}\ }\textbf {\bibinfo {volume}
  {179}},\ \bibinfo {pages} {383 } (\bibinfo {year} {2019})}\BibitemShut
  {NoStop}%
\bibitem [{\citenamefont {Koju}\ and\ \citenamefont
  {Mishin}(2020)}]{Koju2020DirectAM}%
  \BibitemOpen
  \bibfield  {author} {\bibinfo {author} {\bibfnamefont {R.~K.}\ \bibnamefont
  {Koju}}\ and\ \bibinfo {author} {\bibfnamefont {Y.}~\bibnamefont {Mishin}},\
  }\bibfield  {title} {\bibinfo {title} {Direct atomistic modeling of solute
  drag by moving grain boundaries},\ }\href
  {https://doi.org/10.1016/j.actamat.2020.07.052} {\bibfield  {journal}
  {\bibinfo  {journal} {Acta Mater.}\ }\textbf {\bibinfo {volume} {198}},\
  \bibinfo {pages} {111} (\bibinfo {year} {2020})}\BibitemShut {NoStop}%
\bibitem [{\citenamefont {{Darvishi Kamachali}}\ \emph
  {et~al.}(2020)\citenamefont {{Darvishi Kamachali}}, \citenamefont
  {{Kwiatkowski da Silva}}, \citenamefont {McEniry}, \citenamefont {Ponge},
  \citenamefont {Gault}, \citenamefont {Neugebauer},\ and\ \citenamefont
  {Raabe}}]{DarvishiKamachali2020}%
  \BibitemOpen
  \bibfield  {author} {\bibinfo {author} {\bibfnamefont {R.}~\bibnamefont
  {{Darvishi Kamachali}}}, \bibinfo {author} {\bibfnamefont {A.}~\bibnamefont
  {{Kwiatkowski da Silva}}}, \bibinfo {author} {\bibfnamefont {E.}~\bibnamefont
  {McEniry}}, \bibinfo {author} {\bibfnamefont {D.}~\bibnamefont {Ponge}},
  \bibinfo {author} {\bibfnamefont {B.}~\bibnamefont {Gault}}, \bibinfo
  {author} {\bibfnamefont {J.}~\bibnamefont {Neugebauer}},\ and\ \bibinfo
  {author} {\bibfnamefont {D.}~\bibnamefont {Raabe}},\ }\bibfield  {title}
  {\bibinfo {title} {{Segregation-assisted spinodal and transient spinodal
  phase separation at grain boundaries}},\ }\href
  {https://doi.org/10.1038/s41524-020-00456-7} {\bibfield  {journal} {\bibinfo
  {journal} {npj Comput. Mater.}\ }\textbf {\bibinfo {volume} {6}},\ \bibinfo
  {pages} {191} (\bibinfo {year} {2020})}\BibitemShut {NoStop}%
\bibitem [{\citenamefont {Rollett}\ \emph {et~al.}(2015)\citenamefont
  {Rollett}, \citenamefont {Rohrer},\ and\ \citenamefont
  {Suter}}]{Rollett2015}%
  \BibitemOpen
  \bibfield  {author} {\bibinfo {author} {\bibfnamefont {A.}~\bibnamefont
  {Rollett}}, \bibinfo {author} {\bibfnamefont {G.}~\bibnamefont {Rohrer}},\
  and\ \bibinfo {author} {\bibfnamefont {R.}~\bibnamefont {Suter}},\ }\bibfield
   {title} {\bibinfo {title} {{Understanding materials microstructure and
  behavior at the mesoscale}},\ }\href {https://doi.org/10.1557/mrs.2015.262}
  {\bibfield  {journal} {\bibinfo  {journal} {MRS Bulletin}\ }\textbf {\bibinfo
  {volume} {40}},\ \bibinfo {pages} {951} (\bibinfo {year} {2015})}\BibitemShut
  {NoStop}%
\bibitem [{\citenamefont {Elder}\ \emph {et~al.}(2002)\citenamefont {Elder},
  \citenamefont {Katakowski}, \citenamefont {Haataja},\ and\ \citenamefont
  {Grant}}]{Elder2002}%
  \BibitemOpen
  \bibfield  {author} {\bibinfo {author} {\bibfnamefont {K.~R.}\ \bibnamefont
  {Elder}}, \bibinfo {author} {\bibfnamefont {M.}~\bibnamefont {Katakowski}},
  \bibinfo {author} {\bibfnamefont {M.}~\bibnamefont {Haataja}},\ and\ \bibinfo
  {author} {\bibfnamefont {M.}~\bibnamefont {Grant}},\ }\bibfield  {title}
  {\bibinfo {title} {{Modeling Elasticity in Crystal Growth}},\ }\href
  {https://doi.org/10.1103/PhysRevLett.88.245701} {\bibfield  {journal}
  {\bibinfo  {journal} {Phys. Rev. Lett.}\ }\textbf {\bibinfo {volume} {88}},\
  \bibinfo {pages} {245701} (\bibinfo {year} {2002})}\BibitemShut {NoStop}%
\bibitem [{\citenamefont {Elder}\ and\ \citenamefont
  {Grant}(2004)}]{Elder2004}%
  \BibitemOpen
  \bibfield  {author} {\bibinfo {author} {\bibfnamefont {K.~R.}\ \bibnamefont
  {Elder}}\ and\ \bibinfo {author} {\bibfnamefont {M.}~\bibnamefont {Grant}},\
  }\bibfield  {title} {\bibinfo {title} {{Modeling elastic and plastic
  deformations in nonequilibrium processing using phase field crystals}},\
  }\href {https://doi.org/10.1103/PhysRevE.70.051605} {\bibfield  {journal}
  {\bibinfo  {journal} {Phys. Rev. E}\ }\textbf {\bibinfo {volume} {70}},\
  \bibinfo {pages} {051605} (\bibinfo {year} {2004})}\BibitemShut {NoStop}%
\bibitem [{\citenamefont {Elder}\ \emph {et~al.}(2007)\citenamefont {Elder},
  \citenamefont {Provatas}, \citenamefont {Berry}, \citenamefont {Stefanovic},\
  and\ \citenamefont {Grant}}]{Elder2007}%
  \BibitemOpen
  \bibfield  {author} {\bibinfo {author} {\bibfnamefont {K.~R.}\ \bibnamefont
  {Elder}}, \bibinfo {author} {\bibfnamefont {N.}~\bibnamefont {Provatas}},
  \bibinfo {author} {\bibfnamefont {J.}~\bibnamefont {Berry}}, \bibinfo
  {author} {\bibfnamefont {P.}~\bibnamefont {Stefanovic}},\ and\ \bibinfo
  {author} {\bibfnamefont {M.}~\bibnamefont {Grant}},\ }\bibfield  {title}
  {\bibinfo {title} {Phase-field crystal modeling and classical density
  functional theory of freezing},\ }\href
  {https://doi.org/10.1103/PhysRevB.75.064107} {\bibfield  {journal} {\bibinfo
  {journal} {Phys. Rev. B}\ }\textbf {\bibinfo {volume} {75}},\ \bibinfo
  {pages} {064107} (\bibinfo {year} {2007})}\BibitemShut {NoStop}%
\bibitem [{\citenamefont {Skaugen}\ \emph
  {et~al.}(2018{\natexlab{a}})\citenamefont {Skaugen}, \citenamefont
  {Angheluta},\ and\ \citenamefont {Vi\~nals}}]{Skaugen2018b}%
  \BibitemOpen
  \bibfield  {author} {\bibinfo {author} {\bibfnamefont {A.}~\bibnamefont
  {Skaugen}}, \bibinfo {author} {\bibfnamefont {L.}~\bibnamefont {Angheluta}},\
  and\ \bibinfo {author} {\bibfnamefont {J.}~\bibnamefont {Vi\~nals}},\
  }\bibfield  {title} {\bibinfo {title} {Dislocation dynamics and crystal
  plasticity in the phase-field crystal model},\ }\href
  {https://doi.org/10.1103/PhysRevB.97.054113} {\bibfield  {journal} {\bibinfo
  {journal} {Phys. Rev. B}\ }\textbf {\bibinfo {volume} {97}},\ \bibinfo
  {pages} {054113} (\bibinfo {year} {2018}{\natexlab{a}})}\BibitemShut
  {NoStop}%
\bibitem [{\citenamefont {Skaugen}\ \emph
  {et~al.}(2018{\natexlab{b}})\citenamefont {Skaugen}, \citenamefont
  {Angheluta},\ and\ \citenamefont {Vi\~nals}}]{Skaugen2018}%
  \BibitemOpen
  \bibfield  {author} {\bibinfo {author} {\bibfnamefont {A.}~\bibnamefont
  {Skaugen}}, \bibinfo {author} {\bibfnamefont {L.}~\bibnamefont {Angheluta}},\
  and\ \bibinfo {author} {\bibfnamefont {J.}~\bibnamefont {Vi\~nals}},\
  }\bibfield  {title} {\bibinfo {title} {Separation of elastic and plastic
  timescales in a phase field crystal model},\ }\href
  {https://doi.org/10.1103/PhysRevLett.121.255501} {\bibfield  {journal}
  {\bibinfo  {journal} {Phys. Rev. Lett.}\ }\textbf {\bibinfo {volume} {121}},\
  \bibinfo {pages} {255501} (\bibinfo {year} {2018}{\natexlab{b}})}\BibitemShut
  {NoStop}%
\bibitem [{\citenamefont {Salvalaglio}\ \emph {et~al.}(2019)\citenamefont
  {Salvalaglio}, \citenamefont {Voigt},\ and\ \citenamefont
  {Elder}}]{SalvalaglioNPJ2019}%
  \BibitemOpen
  \bibfield  {author} {\bibinfo {author} {\bibfnamefont {M.}~\bibnamefont
  {Salvalaglio}}, \bibinfo {author} {\bibfnamefont {A.}~\bibnamefont {Voigt}},\
  and\ \bibinfo {author} {\bibfnamefont {K.~R.}\ \bibnamefont {Elder}},\
  }\bibfield  {title} {\bibinfo {title} {{Closing the gap between atomic-scale
  lattice deformations and continuum elasticity}},\ }\href
  {https://doi.org/10.1038/s41524-019-0185-0} {\bibfield  {journal} {\bibinfo
  {journal} {npj Comput. Mater.}\ }\textbf {\bibinfo {volume} {5}},\ \bibinfo
  {pages} {48} (\bibinfo {year} {2019})}\BibitemShut {NoStop}%
\bibitem [{\citenamefont {Salvalaglio}\ \emph {et~al.}(2020)\citenamefont
  {Salvalaglio}, \citenamefont {Angheluta}, \citenamefont {Huang},
  \citenamefont {Voigt}, \citenamefont {Elder},\ and\ \citenamefont
  {Viñals}}]{SalvalaglioJMPS2020}%
  \BibitemOpen
  \bibfield  {author} {\bibinfo {author} {\bibfnamefont {M.}~\bibnamefont
  {Salvalaglio}}, \bibinfo {author} {\bibfnamefont {L.}~\bibnamefont
  {Angheluta}}, \bibinfo {author} {\bibfnamefont {Z.-F.}\ \bibnamefont
  {Huang}}, \bibinfo {author} {\bibfnamefont {A.}~\bibnamefont {Voigt}},
  \bibinfo {author} {\bibfnamefont {K.~R.}\ \bibnamefont {Elder}},\ and\
  \bibinfo {author} {\bibfnamefont {J.}~\bibnamefont {Viñals}},\ }\bibfield
  {title} {\bibinfo {title} {A coarse-grained phase-field crystal model of
  plastic motion},\ }\href
  {https://doi.org/https://doi.org/10.1016/j.jmps.2019.103856} {\bibfield
  {journal} {\bibinfo  {journal} {J. Mech. Phys. Solids}\ }\textbf {\bibinfo
  {volume} {137}},\ \bibinfo {pages} {103856} (\bibinfo {year}
  {2020})}\BibitemShut {NoStop}%
\bibitem [{\citenamefont {Goldenfeld}\ \emph {et~al.}(2005)\citenamefont
  {Goldenfeld}, \citenamefont {Athreya},\ and\ \citenamefont
  {Dantzig}}]{Goldenfeld2005}%
  \BibitemOpen
  \bibfield  {author} {\bibinfo {author} {\bibfnamefont {N.}~\bibnamefont
  {Goldenfeld}}, \bibinfo {author} {\bibfnamefont {B.~P.}\ \bibnamefont
  {Athreya}},\ and\ \bibinfo {author} {\bibfnamefont {J.~A.}\ \bibnamefont
  {Dantzig}},\ }\bibfield  {title} {\bibinfo {title} {{Renormalization group
  approach to multiscale simulation of polycrystalline materials using the
  phase field crystal model}},\ }\href
  {https://doi.org/10.1103/PhysRevE.72.020601} {\bibfield  {journal} {\bibinfo
  {journal} {Phys. Rev. E}\ }\textbf {\bibinfo {volume} {72}},\ \bibinfo
  {pages} {020601} (\bibinfo {year} {2005})}\BibitemShut {NoStop}%
\bibitem [{\citenamefont {Athreya}\ \emph {et~al.}(2006)\citenamefont
  {Athreya}, \citenamefont {Goldenfeld},\ and\ \citenamefont
  {Dantzig}}]{Athreya2006}%
  \BibitemOpen
  \bibfield  {author} {\bibinfo {author} {\bibfnamefont {B.~P.}\ \bibnamefont
  {Athreya}}, \bibinfo {author} {\bibfnamefont {N.}~\bibnamefont
  {Goldenfeld}},\ and\ \bibinfo {author} {\bibfnamefont {J.~A.}\ \bibnamefont
  {Dantzig}},\ }\bibfield  {title} {\bibinfo {title} {{Renormalization-group
  theory for the phase-field crystal equation}},\ }\href
  {https://doi.org/10.1103/PhysRevE.74.011601} {\bibfield  {journal} {\bibinfo
  {journal} {Phys. Rev. E}\ }\textbf {\bibinfo {volume} {74}},\ \bibinfo
  {pages} {011601} (\bibinfo {year} {2006})}\BibitemShut {NoStop}%
\bibitem [{\citenamefont {Elder}\ \emph {et~al.}(2010)\citenamefont {Elder},
  \citenamefont {Huang},\ and\ \citenamefont {Provatas}}]{ElderPRE2010}%
  \BibitemOpen
  \bibfield  {author} {\bibinfo {author} {\bibfnamefont {K.~R.}\ \bibnamefont
  {Elder}}, \bibinfo {author} {\bibfnamefont {Z.-F.}\ \bibnamefont {Huang}},\
  and\ \bibinfo {author} {\bibfnamefont {N.}~\bibnamefont {Provatas}},\
  }\bibfield  {title} {\bibinfo {title} {{Amplitude expansion of the binary
  phase-field-crystal model}},\ }\href
  {https://doi.org/10.1103/PhysRevE.81.011602} {\bibfield  {journal} {\bibinfo
  {journal} {Phys. Rev. E}\ }\textbf {\bibinfo {volume} {81}},\ \bibinfo
  {pages} {011602} (\bibinfo {year} {2010})}\BibitemShut {NoStop}%
\bibitem [{\citenamefont {Huang}\ \emph {et~al.}(2010)\citenamefont {Huang},
  \citenamefont {Elder},\ and\ \citenamefont {Provatas}}]{Huang2010}%
  \BibitemOpen
  \bibfield  {author} {\bibinfo {author} {\bibfnamefont {Z.-F.}\ \bibnamefont
  {Huang}}, \bibinfo {author} {\bibfnamefont {K.~R.}\ \bibnamefont {Elder}},\
  and\ \bibinfo {author} {\bibfnamefont {N.}~\bibnamefont {Provatas}},\
  }\bibfield  {title} {\bibinfo {title} {Phase-field-crystal dynamics for
  binary systems: Derivation from dynamical density functional theory,
  amplitude equation formalism, and applications to alloy heterostructures},\
  }\href {https://doi.org/10.1103/PhysRevE.82.021605} {\bibfield  {journal}
  {\bibinfo  {journal} {Phys. Rev. E}\ }\textbf {\bibinfo {volume} {82}},\
  \bibinfo {pages} {021605} (\bibinfo {year} {2010})}\BibitemShut {NoStop}%
\bibitem [{fre()}]{free}%
  \BibitemOpen
  \href@noop {} {}\bibinfo {note} {In the original binary PFC model
  \cite{Elder2007} the free energy contains terms $n(1+\nabla^2)^2n$ and $\psi
  n \nabla^2(1+\nabla^2)n$ which transform in the amplitude formulation to
  $2\sum |{\cal G}_n\eta_n|^2$ and $-\sum(\eta_n{\cal G}_n^*\eta_n^*+{\rm
  c.c.})\psi$ for 2D triangular and 3D bcc systems, requiring modes of only one
  length scale ($|\vec{q}_n|=1$). For 2D square and 3D fcc lattices two modes
  are needed and it is convenient to use
  $n[q_1^2/(q_1^2-1)]^2(1+\Lap)^2(1+\Lap/q_1^2)^2n$ and
  $[q_1^2/(q_1^2-1)]^2\psi n
  \nabla^2(1+\nabla^2)(1+\Lap/q_1^2)(q_1^2+1+2\Lap)$, which then transform to
  $2\sum |{\cal G}_n\eta_n|^2$ and $-\sum q_n^2(\eta_n{\cal G}_n^*\eta_n^*+{\rm
  c.c.})\psi$ respectively, where $q_1$ corresponds to the other length
  scale.}\BibitemShut {Stop}%
\bibitem [{Sup()}]{Suppl}%
  \BibitemOpen
  \href@noop {} {}\bibinfo {note} {See Supplemental Material for the
  discretization scheme and the setups of initial conditions for numerical
  simulations, the expressions of dislocation velocity for various crystalline
  symmetries, the expressions of the compositional contribution to strain and
  stress for binary alloys, and some videos of dislocation motion. It includes Ref.~\cite{cai2006non}.}\BibitemShut
  {Stop}%
\bibitem [{\citenamefont {Huang}(2013)}]{Huang2013}%
  \BibitemOpen
  \bibfield  {author} {\bibinfo {author} {\bibfnamefont {Z.-F.}\ \bibnamefont
  {Huang}},\ }\bibfield  {title} {\bibinfo {title} {Scale-coupling and
  interface-pinning effects in the phase field crystal model},\ }\href
  {https://doi.org/10.1103/PhysRevE.87.012401} {\bibfield  {journal} {\bibinfo
  {journal} {Phys. Rev. E}\ }\textbf {\bibinfo {volume} {87}},\ \bibinfo
  {pages} {012401} (\bibinfo {year} {2013})}\BibitemShut {NoStop}%
\bibitem [{\citenamefont {Huang}(2016)}]{Huang2016}%
  \BibitemOpen
  \bibfield  {author} {\bibinfo {author} {\bibfnamefont {Z.-F.}\ \bibnamefont
  {Huang}},\ }\bibfield  {title} {\bibinfo {title} {Scaling of alloy
  interfacial properties under compositional strain},\ }\href
  {https://doi.org/10.1103/PhysRevE.93.022803} {\bibfield  {journal} {\bibinfo
  {journal} {Phys. Rev. E}\ }\textbf {\bibinfo {volume} {93}},\ \bibinfo
  {pages} {022803} (\bibinfo {year} {2016})}\BibitemShut {NoStop}%
\bibitem [{\citenamefont {Heinonen}\ \emph {et~al.}(2014)\citenamefont
  {Heinonen}, \citenamefont {Achim}, \citenamefont {Elder}, \citenamefont
  {Buyukdagli},\ and\ \citenamefont {Ala-Nissila}}]{Heinonen2014}%
  \BibitemOpen
  \bibfield  {author} {\bibinfo {author} {\bibfnamefont {V.}~\bibnamefont
  {Heinonen}}, \bibinfo {author} {\bibfnamefont {C.~V.}\ \bibnamefont {Achim}},
  \bibinfo {author} {\bibfnamefont {K.~R.}\ \bibnamefont {Elder}}, \bibinfo
  {author} {\bibfnamefont {S.}~\bibnamefont {Buyukdagli}},\ and\ \bibinfo
  {author} {\bibfnamefont {T.}~\bibnamefont {Ala-Nissila}},\ }\bibfield
  {title} {\bibinfo {title} {{Phase-field-crystal models and mechanical
  equilibrium}},\ }\href {https://doi.org/10.1103/PhysRevE.89.032411}
  {\bibfield  {journal} {\bibinfo  {journal} {Phys. Rev. E}\ }\textbf {\bibinfo
  {volume} {89}},\ \bibinfo {pages} {032411} (\bibinfo {year}
  {2014})}\BibitemShut {NoStop}%
\bibitem [{\citenamefont {Lubarda}(2019)}]{pk2019}%
  \BibitemOpen
  \bibfield  {author} {\bibinfo {author} {\bibfnamefont {V.~A.}\ \bibnamefont
  {Lubarda}},\ }\bibfield  {title} {\bibinfo {title} {Dislocation {Burgers}
  vector and the {Peach–Koehler} force: a review},\ }\href
  {https://doi.org/10.1016/j.jmrt.2018.08.014} {\bibfield  {journal} {\bibinfo
  {journal} {J. Matter. Res. Technol.}\ }\textbf {\bibinfo {volume} {8}},\
  \bibinfo {pages} {1550} (\bibinfo {year} {2019})}\BibitemShut {NoStop}%
\bibitem [{\citenamefont {Landau}\ and\ \citenamefont {Lifshitz}(1970)}]{LLEL}%
  \BibitemOpen
  \bibfield  {author} {\bibinfo {author} {\bibfnamefont {L.~D.}\ \bibnamefont
  {Landau}}\ and\ \bibinfo {author} {\bibfnamefont {E.~M.}\ \bibnamefont
  {Lifshitz}},\ }\href@noop {} {\emph {\bibinfo {title} {Theory of
  Elasticity}}},\ \bibinfo {edition} {2nd}\ ed.\ (\bibinfo  {publisher}
  {Peramon Press Ltd.},\ \bibinfo {address} {Oxford, England},\ \bibinfo {year}
  {1970})\BibitemShut {NoStop}%
\bibitem [{\citenamefont {Vey}\ and\ \citenamefont {Voigt}(2007)}]{Vey2007}%
  \BibitemOpen
  \bibfield  {author} {\bibinfo {author} {\bibfnamefont {S.}~\bibnamefont
  {Vey}}\ and\ \bibinfo {author} {\bibfnamefont {A.}~\bibnamefont {Voigt}},\
  }\bibfield  {title} {\bibinfo {title} {{AMDiS}: adaptive multidimensional
  simulations},\ }\href {https://doi.org/10.1007/s00791-006-0048-3} {\bibfield
  {journal} {\bibinfo  {journal} {Comput. Visual. Sci.}\ }\textbf {\bibinfo
  {volume} {10}},\ \bibinfo {pages} {57} (\bibinfo {year} {2007})}\BibitemShut
  {NoStop}%
\bibitem [{\citenamefont {Witkowski}\ \emph {et~al.}(2015)\citenamefont
  {Witkowski}, \citenamefont {Ling}, \citenamefont {Praetorius},\ and\
  \citenamefont {Voigt}}]{Witkowski2015}%
  \BibitemOpen
  \bibfield  {author} {\bibinfo {author} {\bibfnamefont {T.}~\bibnamefont
  {Witkowski}}, \bibinfo {author} {\bibfnamefont {S.}~\bibnamefont {Ling}},
  \bibinfo {author} {\bibfnamefont {S.}~\bibnamefont {Praetorius}},\ and\
  \bibinfo {author} {\bibfnamefont {A.}~\bibnamefont {Voigt}},\ }\bibfield
  {title} {\bibinfo {title} {Software concepts and numerical algorithms for a
  scalable adaptive parallel finite element method},\ }\href
  {https://doi.org/10.1007/s10444-015-9405-4} {\bibfield  {journal} {\bibinfo
  {journal} {Adv. Comput. Math.}\ }\textbf {\bibinfo {volume} {41}},\ \bibinfo
  {pages} {1145} (\bibinfo {year} {2015})}\BibitemShut {NoStop}%
\bibitem [{\citenamefont {Salvalaglio}\ \emph {et~al.}(2017)\citenamefont
  {Salvalaglio}, \citenamefont {Backofen}, \citenamefont {Voigt},\ and\
  \citenamefont {Elder}}]{SalvalaglioAPFC2017}%
  \BibitemOpen
  \bibfield  {author} {\bibinfo {author} {\bibfnamefont {M.}~\bibnamefont
  {Salvalaglio}}, \bibinfo {author} {\bibfnamefont {R.}~\bibnamefont
  {Backofen}}, \bibinfo {author} {\bibfnamefont {A.}~\bibnamefont {Voigt}},\
  and\ \bibinfo {author} {\bibfnamefont {K.~R.}\ \bibnamefont {Elder}},\
  }\bibfield  {title} {\bibinfo {title} {{Controlling the energy of defects and
  interfaces in the amplitude expansion of the phase-field crystal model}},\
  }\href {https://doi.org/10.1103/PhysRevE.96.023301} {\bibfield  {journal}
  {\bibinfo  {journal} {Phys. Rev. E}\ }\textbf {\bibinfo {volume} {96}},\
  \bibinfo {pages} {023301} (\bibinfo {year} {2017})}\BibitemShut {NoStop}%
\bibitem [{\citenamefont {Praetorius}\ \emph {et~al.}(2019)\citenamefont
  {Praetorius}, \citenamefont {Salvalaglio},\ and\ \citenamefont
  {Voigt}}]{Praetorius_2019}%
  \BibitemOpen
  \bibfield  {author} {\bibinfo {author} {\bibfnamefont {S.}~\bibnamefont
  {Praetorius}}, \bibinfo {author} {\bibfnamefont {M.}~\bibnamefont
  {Salvalaglio}},\ and\ \bibinfo {author} {\bibfnamefont {A.}~\bibnamefont
  {Voigt}},\ }\bibfield  {title} {\bibinfo {title} {An efficient numerical
  framework for the amplitude expansion of the phase-field crystal model},\
  }\href {https://doi.org/10.1088/1361-651x/ab1508} {\bibfield  {journal}
  {\bibinfo  {journal} {Model. Simul. Mater. Sci. Eng.}\ }\textbf {\bibinfo
  {volume} {27}},\ \bibinfo {pages} {044004} (\bibinfo {year}
  {2019})}\BibitemShut {NoStop}%
\bibitem [{\citenamefont {H\"uter}\ \emph {et~al.}(2016)\citenamefont
  {H\"uter}, \citenamefont {Fri\'ak}, \citenamefont {Weikamp}, \citenamefont
  {Neugebauer}, \citenamefont {Goldenfeld}, \citenamefont {Svendsen},\ and\
  \citenamefont {Spatschek}}]{Huter2016}%
  \BibitemOpen
  \bibfield  {author} {\bibinfo {author} {\bibfnamefont {C.}~\bibnamefont
  {H\"uter}}, \bibinfo {author} {\bibfnamefont {M.}~\bibnamefont {Fri\'ak}},
  \bibinfo {author} {\bibfnamefont {M.}~\bibnamefont {Weikamp}}, \bibinfo
  {author} {\bibfnamefont {J.}~\bibnamefont {Neugebauer}}, \bibinfo {author}
  {\bibfnamefont {N.}~\bibnamefont {Goldenfeld}}, \bibinfo {author}
  {\bibfnamefont {B.}~\bibnamefont {Svendsen}},\ and\ \bibinfo {author}
  {\bibfnamefont {R.}~\bibnamefont {Spatschek}},\ }\bibfield  {title} {\bibinfo
  {title} {Nonlinear elastic effects in phase field crystal and amplitude
  equations: Comparison to ab initio simulations of bcc metals and graphene},\
  }\href {https://doi.org/10.1103/PhysRevB.93.214105} {\bibfield  {journal}
  {\bibinfo  {journal} {Phys. Rev. B}\ }\textbf {\bibinfo {volume} {93}},\
  \bibinfo {pages} {214105} (\bibinfo {year} {2016})}\BibitemShut {NoStop}%
\bibitem [{\citenamefont {Anderson}\ \emph {et~al.}(2017)\citenamefont
  {Anderson}, \citenamefont {Hirth},\ and\ \citenamefont
  {Lothe}}]{anderson2017}%
  \BibitemOpen
  \bibfield  {author} {\bibinfo {author} {\bibfnamefont {P.}~\bibnamefont
  {Anderson}}, \bibinfo {author} {\bibfnamefont {J.}~\bibnamefont {Hirth}},\
  and\ \bibinfo {author} {\bibfnamefont {J.}~\bibnamefont {Lothe}},\
  }\href@noop {} {\emph {\bibinfo {title} {Theory of Dislocations}}}\ (\bibinfo
   {publisher} {Cambridge University Press},\ \bibinfo {year}
  {2017})\BibitemShut {NoStop}%
\bibitem [{\citenamefont {Salvalaglio}\ \emph {et~al.}(2018)\citenamefont
  {Salvalaglio}, \citenamefont {Backofen}, \citenamefont {Elder},\ and\
  \citenamefont {Voigt}}]{SalvalaglioPRM2018}%
  \BibitemOpen
  \bibfield  {author} {\bibinfo {author} {\bibfnamefont {M.}~\bibnamefont
  {Salvalaglio}}, \bibinfo {author} {\bibfnamefont {R.}~\bibnamefont
  {Backofen}}, \bibinfo {author} {\bibfnamefont {K.~R.}\ \bibnamefont
  {Elder}},\ and\ \bibinfo {author} {\bibfnamefont {A.}~\bibnamefont {Voigt}},\
  }\bibfield  {title} {\bibinfo {title} {{Defects at grain boundaries: A
  coarse-grained, three-dimensional description by the amplitude expansion of
  the phase-field crystal model}},\ }\href
  {https://doi.org/10.1103/PhysRevMaterials.2.053804} {\bibfield  {journal}
  {\bibinfo  {journal} {Phys. Rev. Materials}\ }\textbf {\bibinfo {volume}
  {2}},\ \bibinfo {pages} {053804} (\bibinfo {year} {2018})}\BibitemShut
  {NoStop}%
\bibitem [{\citenamefont {Doherty}\ \emph {et~al.}(1997)\citenamefont
  {Doherty}, \citenamefont {Hughes}, \citenamefont {Humphreys}, \citenamefont
  {Jonas}, \citenamefont {Jensen}, \citenamefont {Kassner}, \citenamefont
  {King}, \citenamefont {McNelley}, \citenamefont {McQueen},\ and\
  \citenamefont {Rollett}}]{Doherty1997}%
  \BibitemOpen
  \bibfield  {author} {\bibinfo {author} {\bibfnamefont {R.}~\bibnamefont
  {Doherty}}, \bibinfo {author} {\bibfnamefont {D.}~\bibnamefont {Hughes}},
  \bibinfo {author} {\bibfnamefont {F.}~\bibnamefont {Humphreys}}, \bibinfo
  {author} {\bibfnamefont {J.}~\bibnamefont {Jonas}}, \bibinfo {author}
  {\bibfnamefont {D.}~\bibnamefont {Jensen}}, \bibinfo {author} {\bibfnamefont
  {M.}~\bibnamefont {Kassner}}, \bibinfo {author} {\bibfnamefont
  {W.}~\bibnamefont {King}}, \bibinfo {author} {\bibfnamefont {T.}~\bibnamefont
  {McNelley}}, \bibinfo {author} {\bibfnamefont {H.}~\bibnamefont {McQueen}},\
  and\ \bibinfo {author} {\bibfnamefont {A.}~\bibnamefont {Rollett}},\
  }\bibfield  {title} {\bibinfo {title} {{Current issues in recrystallization:
  a review}},\ }\href {https://doi.org/10.1016/S0921-5093(97)00424-3}
  {\bibfield  {journal} {\bibinfo  {journal} {Materials Science and
  Engineering: A}\ }\textbf {\bibinfo {volume} {238}},\ \bibinfo {pages} {219}
  (\bibinfo {year} {1997})}\BibitemShut {NoStop}%
\bibitem [{\citenamefont {Yamanaka}\ \emph {et~al.}(2017)\citenamefont
  {Yamanaka}, \citenamefont {McReynolds},\ and\ \citenamefont
  {Voorhees}}]{Yamanaka2017}%
  \BibitemOpen
  \bibfield  {author} {\bibinfo {author} {\bibfnamefont {A.}~\bibnamefont
  {Yamanaka}}, \bibinfo {author} {\bibfnamefont {K.}~\bibnamefont
  {McReynolds}},\ and\ \bibinfo {author} {\bibfnamefont {P.~W.}\ \bibnamefont
  {Voorhees}},\ }\bibfield  {title} {\bibinfo {title} {{Phase field crystal
  simulation of grain boundary motion, grain rotation and dislocation reactions
  in a BCC bicrystal}},\ }\href {https://doi.org/10.1016/j.actamat.2017.05.022}
  {\bibfield  {journal} {\bibinfo  {journal} {Acta Mater.}\ }\textbf {\bibinfo
  {volume} {133}},\ \bibinfo {pages} {160} (\bibinfo {year}
  {2017})}\BibitemShut {NoStop}%
\bibitem [{\citenamefont {Nye}(1953)}]{nye1953some}%
  \BibitemOpen
  \bibfield  {author} {\bibinfo {author} {\bibfnamefont {J.}~\bibnamefont
  {Nye}},\ }\bibfield  {title} {\bibinfo {title} {Some geometrical relations in
  dislocated crystals},\ }\href {https://doi.org/10.1016/0001-6160(53)90054-6}
  {\bibfield  {journal} {\bibinfo  {journal} {Acta Metall.}\ }\textbf {\bibinfo
  {volume} {1}},\ \bibinfo {pages} {153} (\bibinfo {year} {1953})}\BibitemShut
  {NoStop}%
\bibitem [{\citenamefont {Hartley}\ and\ \citenamefont
  {Mishin}(2005)}]{hartley2005characterization}%
  \BibitemOpen
  \bibfield  {author} {\bibinfo {author} {\bibfnamefont {C.}~\bibnamefont
  {Hartley}}\ and\ \bibinfo {author} {\bibfnamefont {Y.}~\bibnamefont
  {Mishin}},\ }\bibfield  {title} {\bibinfo {title} {Characterization and
  visualization of the lattice misfit associated with dislocation cores},\
  }\href {https://doi.org/10.1016/j.actamat.2004.11.027} {\bibfield  {journal}
  {\bibinfo  {journal} {Acta Mater.}\ }\textbf {\bibinfo {volume} {53}},\
  \bibinfo {pages} {1313} (\bibinfo {year} {2005})}\BibitemShut {NoStop}%
  \bibitem [{\citenamefont {Cai}\ \emph {et~al.}(2006)\citenamefont {Cai},
  \citenamefont {Arsenlis}, \citenamefont {Weinberger},\ and\ \citenamefont
  {Bulatov}}]{cai2006non}%
  \BibitemOpen
  \bibfield  {author} {\bibinfo {author} {\bibfnamefont {W.}~\bibnamefont
  {Cai}}, \bibinfo {author} {\bibfnamefont {A.}~\bibnamefont {Arsenlis}},
  \bibinfo {author} {\bibfnamefont {C.~R.}\ \bibnamefont {Weinberger}},\ and\
  \bibinfo {author} {\bibfnamefont {V.~V.}\ \bibnamefont {Bulatov}},\
  }\bibfield  {title} {\bibinfo {title} {A non-singular continuum theory of
  dislocations},\ }\href {https://doi.org/10.1016/j.jmps.2005.09.005}
  {\bibfield  {journal} {\bibinfo  {journal} {J. Mech. Phys. Solids}\ }\textbf
  {\bibinfo {volume} {54}},\ \bibinfo {pages} {561} (\bibinfo {year}
  {2006})}\BibitemShut {NoStop}%
\end{thebibliography}

\begin{thebibliography}{10}%
\makeatletter
\providecommand \@ifxundefined [1]{%
 \@ifx{#1\undefined}
}%
\providecommand \@ifnum [1]{%
 \ifnum #1\expandafter \@firstoftwo
 \else \expandafter \@secondoftwo
 \fi
}%
\providecommand \@ifx [1]{%
 \ifx #1\expandafter \@firstoftwo
 \else \expandafter \@secondoftwo
 \fi
}%
\providecommand \natexlab [1]{#1}%
\providecommand \enquote  [1]{``#1''}%
\providecommand \bibnamefont  [1]{#1}%
\providecommand \bibfnamefont [1]{#1}%
\providecommand \citenamefont [1]{#1}%
\providecommand \href@noop [0]{\@secondoftwo}%
\providecommand \href [0]{\begingroup \@sanitize@url \@href}%
\providecommand \@href[1]{\@@startlink{#1}\@@href}%
\providecommand \@@href[1]{\endgroup#1\@@endlink}%
\providecommand \@sanitize@url [0]{\catcode `\\12\catcode `\$12\catcode
  `\&12\catcode `\#12\catcode `\^12\catcode `\_12\catcode `\%12\relax}%
\providecommand \@@startlink[1]{}%
\providecommand \@@endlink[0]{}%
\providecommand \url  [0]{\begingroup\@sanitize@url \@url }%
\providecommand \@url [1]{\endgroup\@href {#1}{\urlprefix }}%
\providecommand \urlprefix  [0]{URL }%
\providecommand \Eprint [0]{\href }%
\providecommand \doibase [0]{https://doi.org/}%
\providecommand \selectlanguage [0]{\@gobble}%
\providecommand \bibinfo  [0]{\@secondoftwo}%
\providecommand \bibfield  [0]{\@secondoftwo}%
\providecommand \translation [1]{[#1]}%
\providecommand \BibitemOpen [0]{}%
\providecommand \bibitemStop [0]{}%
\providecommand \bibitemNoStop [0]{.\EOS\space}%
\providecommand \EOS [0]{\spacefactor3000\relax}%
\providecommand \BibitemShut  [1]{\csname bibitem#1\endcsname}%
\let\auto@bib@innerbib\@empty
\bibitem [{\citenamefont {Elder}\ \emph {et~al.}(2010)\citenamefont {Elder},
  \citenamefont {Huang},\ and\ \citenamefont {Provatas}}]{SMElderPRE2010}%
  \BibitemOpen
  \bibfield  {author} {\bibinfo {author} {\bibfnamefont {K.~R.}\ \bibnamefont
  {Elder}}, \bibinfo {author} {\bibfnamefont {Z.-F.}\ \bibnamefont {Huang}},\
  and\ \bibinfo {author} {\bibfnamefont {N.}~\bibnamefont {Provatas}},\
  }\bibfield  {title} {\bibinfo {title} {{Amplitude expansion of the binary
  phase-field-crystal model}},\ }\href
  {https://doi.org/10.1103/PhysRevE.81.011602} {\bibfield  {journal} {\bibinfo
  {journal} {Phys. Rev. E}\ }\textbf {\bibinfo {volume} {81}},\ \bibinfo
  {pages} {011602} (\bibinfo {year} {2010})}\BibitemShut {NoStop}%
\bibitem [{\citenamefont {Salvalaglio}\ \emph {et~al.}(2017)\citenamefont
  {Salvalaglio}, \citenamefont {Backofen}, \citenamefont {Voigt},\ and\
  \citenamefont {Elder}}]{SMSalvalaglioAPFC2017}%
  \BibitemOpen
  \bibfield  {author} {\bibinfo {author} {\bibfnamefont {M.}~\bibnamefont
  {Salvalaglio}}, \bibinfo {author} {\bibfnamefont {R.}~\bibnamefont
  {Backofen}}, \bibinfo {author} {\bibfnamefont {A.}~\bibnamefont {Voigt}},\
  and\ \bibinfo {author} {\bibfnamefont {K.~R.}\ \bibnamefont {Elder}},\
  }\bibfield  {title} {\bibinfo {title} {{Controlling the energy of defects and
  interfaces in the amplitude expansion of the phase-field crystal model}},\
  }\href {https://doi.org/10.1103/PhysRevE.96.023301} {\bibfield  {journal}
  {\bibinfo  {journal} {Phys. Rev. E}\ }\textbf {\bibinfo {volume} {96}},\
  \bibinfo {pages} {023301} (\bibinfo {year} {2017})}\BibitemShut {NoStop}%
\bibitem [{\citenamefont {Praetorius}\ \emph {et~al.}(2019)\citenamefont
  {Praetorius}, \citenamefont {Salvalaglio},\ and\ \citenamefont
  {Voigt}}]{SMPraetorius_2019}%
  \BibitemOpen
  \bibfield  {author} {\bibinfo {author} {\bibfnamefont {S.}~\bibnamefont
  {Praetorius}}, \bibinfo {author} {\bibfnamefont {M.}~\bibnamefont
  {Salvalaglio}},\ and\ \bibinfo {author} {\bibfnamefont {A.}~\bibnamefont
  {Voigt}},\ }\bibfield  {title} {\bibinfo {title} {An efficient numerical
  framework for the amplitude expansion of the phase-field crystal model},\
  }\href {https://doi.org/10.1088/1361-651x/ab1508} {\bibfield  {journal}
  {\bibinfo  {journal} {Model. Simul. Mater. Sci. Eng.}\ }\textbf {\bibinfo
  {volume} {27}},\ \bibinfo {pages} {044004} (\bibinfo {year}
  {2019})}\BibitemShut {NoStop}%
\bibitem [{\citenamefont {Vey}\ and\ \citenamefont {Voigt}(2007)}]{SMVey2007}%
  \BibitemOpen
  \bibfield  {author} {\bibinfo {author} {\bibfnamefont {S.}~\bibnamefont
  {Vey}}\ and\ \bibinfo {author} {\bibfnamefont {A.}~\bibnamefont {Voigt}},\
  }\bibfield  {title} {\bibinfo {title} {{AMDiS}: adaptive multidimensional
  simulations},\ }\href {https://doi.org/10.1007/s00791-006-0048-3} {\bibfield
  {journal} {\bibinfo  {journal} {Comput. Visual. Sci.}\ }\textbf {\bibinfo
  {volume} {10}},\ \bibinfo {pages} {57} (\bibinfo {year} {2007})}\BibitemShut
  {NoStop}%
\bibitem [{\citenamefont {Witkowski}\ \emph {et~al.}(2015)\citenamefont
  {Witkowski}, \citenamefont {Ling}, \citenamefont {Praetorius},\ and\
  \citenamefont {Voigt}}]{SMWitkowski2015}%
  \BibitemOpen
  \bibfield  {author} {\bibinfo {author} {\bibfnamefont {T.}~\bibnamefont
  {Witkowski}}, \bibinfo {author} {\bibfnamefont {S.}~\bibnamefont {Ling}},
  \bibinfo {author} {\bibfnamefont {S.}~\bibnamefont {Praetorius}},\ and\
  \bibinfo {author} {\bibfnamefont {A.}~\bibnamefont {Voigt}},\ }\bibfield
  {title} {\bibinfo {title} {Software concepts and numerical algorithms for a
  scalable adaptive parallel finite element method},\ }\href
  {https://doi.org/10.1007/s10444-015-9405-4} {\bibfield  {journal} {\bibinfo
  {journal} {Adv. Comput. Math.}\ }\textbf {\bibinfo {volume} {41}},\ \bibinfo
  {pages} {1145} (\bibinfo {year} {2015})}\BibitemShut {NoStop}%
\bibitem [{\citenamefont {Huang}\ \emph {et~al.}(2010)\citenamefont {Huang},
  \citenamefont {Elder},\ and\ \citenamefont {Provatas}}]{SMHuang10}%
  \BibitemOpen
  \bibfield  {author} {\bibinfo {author} {\bibfnamefont {Z.-F.}\ \bibnamefont
  {Huang}}, \bibinfo {author} {\bibfnamefont {K.~R.}\ \bibnamefont {Elder}},\
  and\ \bibinfo {author} {\bibfnamefont {N.}~\bibnamefont {Provatas}},\
  }\bibfield  {title} {\bibinfo {title} {Phase-field-crystal dynamics for
  binary systems: Derivation from dynamical density functional theory,
  amplitude equation formalism, and applications to alloy heterostructures},\
  }\href {https://doi.org/10.1103/PhysRevE.82.021605} {\bibfield  {journal}
  {\bibinfo  {journal} {Phys. Rev. E}\ }\textbf {\bibinfo {volume} {82}},\
  \bibinfo {pages} {021605} (\bibinfo {year} {2010})}\BibitemShut {NoStop}%
\bibitem [{\citenamefont {Anderson}\ \emph {et~al.}(2017)\citenamefont
  {Anderson}, \citenamefont {Hirth},\ and\ \citenamefont
  {Lothe}}]{SManderson2017}%
  \BibitemOpen
  \bibfield  {author} {\bibinfo {author} {\bibfnamefont {P.}~\bibnamefont
  {Anderson}}, \bibinfo {author} {\bibfnamefont {J.}~\bibnamefont {Hirth}},\
  and\ \bibinfo {author} {\bibfnamefont {J.}~\bibnamefont {Lothe}},\
  }\href@noop {} {\emph {\bibinfo {title} {Theory of Dislocations}}}\ (\bibinfo
   {publisher} {Cambridge University Press},\ \bibinfo {year}
  {2017})\BibitemShut {NoStop}%
\bibitem [{\citenamefont {Cai}\ \emph {et~al.}(2006)\citenamefont {Cai},
  \citenamefont {Arsenlis}, \citenamefont {Weinberger},\ and\ \citenamefont
  {Bulatov}}]{SMcai2006non}%
  \BibitemOpen
  \bibfield  {author} {\bibinfo {author} {\bibfnamefont {W.}~\bibnamefont
  {Cai}}, \bibinfo {author} {\bibfnamefont {A.}~\bibnamefont {Arsenlis}},
  \bibinfo {author} {\bibfnamefont {C.~R.}\ \bibnamefont {Weinberger}},\ and\
  \bibinfo {author} {\bibfnamefont {V.~V.}\ \bibnamefont {Bulatov}},\
  }\bibfield  {title} {\bibinfo {title} {A non-singular continuum theory of
  dislocations},\ }\href {https://doi.org/10.1016/j.jmps.2005.09.005}
  {\bibfield  {journal} {\bibinfo  {journal} {J. Mech. Phys. Solids}\ }\textbf
  {\bibinfo {volume} {54}},\ \bibinfo {pages} {561} (\bibinfo {year}
  {2006})}\BibitemShut {NoStop}%
\bibitem [{\citenamefont {Salvalaglio}\ \emph {et~al.}(2019)\citenamefont
  {Salvalaglio}, \citenamefont {Voigt},\ and\ \citenamefont
  {Elder}}]{SMSalvalaglioNPJ2019}%
  \BibitemOpen
  \bibfield  {author} {\bibinfo {author} {\bibfnamefont {M.}~\bibnamefont
  {Salvalaglio}}, \bibinfo {author} {\bibfnamefont {A.}~\bibnamefont {Voigt}},\
  and\ \bibinfo {author} {\bibfnamefont {K.~R.}\ \bibnamefont {Elder}},\
  }\bibfield  {title} {\bibinfo {title} {{Closing the gap between atomic-scale
  lattice deformations and continuum elasticity}},\ }\href
  {https://doi.org/10.1038/s41524-019-0185-0} {\bibfield  {journal} {\bibinfo
  {journal} {npj Comput. Mater.}\ }\textbf {\bibinfo {volume} {5}},\ \bibinfo
  {pages} {48} (\bibinfo {year} {2019})}\BibitemShut {NoStop}%
\bibitem [{\citenamefont {Salvalaglio}\ \emph {et~al.}(2020)\citenamefont
  {Salvalaglio}, \citenamefont {Angheluta}, \citenamefont {Huang},
  \citenamefont {Voigt}, \citenamefont {Elder},\ and\ \citenamefont
  {Viñals}}]{SMSalvalaglioJMPS2020}%
  \BibitemOpen
  \bibfield  {author} {\bibinfo {author} {\bibfnamefont {M.}~\bibnamefont
  {Salvalaglio}}, \bibinfo {author} {\bibfnamefont {L.}~\bibnamefont
  {Angheluta}}, \bibinfo {author} {\bibfnamefont {Z.-F.}\ \bibnamefont
  {Huang}}, \bibinfo {author} {\bibfnamefont {A.}~\bibnamefont {Voigt}},
  \bibinfo {author} {\bibfnamefont {K.~R.}\ \bibnamefont {Elder}},\ and\
  \bibinfo {author} {\bibfnamefont {J.}~\bibnamefont {Viñals}},\ }\bibfield
  {title} {\bibinfo {title} {A coarse-grained phase-field crystal model of
  plastic motion},\ }\href
  {https://doi.org/https://doi.org/10.1016/j.jmps.2019.103856} {\bibfield
  {journal} {\bibinfo  {journal} {J. Mech. Phys. Solids}\ }\textbf {\bibinfo
  {volume} {137}},\ \bibinfo {pages} {103856} (\bibinfo {year}
  {2020})}\BibitemShut {NoStop}%
\end{thebibliography}
\end{document}